\documentclass[twocolumn,superscriptaddress,showpacs]{revtex4}

\usepackage[bookmarks=false]{hyperref}
\usepackage[english]{babel}
\usepackage[utf8]{inputenc}
\usepackage{amsmath}
\usepackage{amssymb}
\usepackage{graphicx}

\begin{document}

\title{Electron Weibel Instability in Relativistic Counter-Streaming Plasmas with Flow-Aligned External Magnetic Fields}

\author{A. Grassi}
\email[Corresponding author: ]{a.grassi8@gmail.com }
\affiliation{LULI, UPMC Université Paris 06: Sorbonne Universités, CNRS, Ecole Polytechnique, CEA, Université Paris-Saclay, F-75252 Paris Cedex 05, France}
\affiliation{Dipartimento di Fisica Enrico Fermi, Università di Pisa, Largo Bruno Pontecorvo 3, I-56127 Pisa, Italy}
\affiliation{Istituto Nazionale di Ottica, Consiglio Nazionale delle Ricerche (CNR/INO), u.o.s. Adriano Gozzini, I-56127 Pisa, Italy}
\author{M. Grech}
\affiliation{LULI, CNRS, Ecole Polytechnique, CEA, Université Paris-Saclay, UPMC Université Paris 06: Sorbonne Universités, F-91128 Palaiseau Cedex, France}
\author{F. Amiranoff}
\affiliation{LULI, CNRS, Ecole Polytechnique, CEA, Université Paris-Saclay, UPMC Université Paris 06: Sorbonne Universités, F-91128 Palaiseau Cedex, France}
\author{F. Pegoraro}
\affiliation{Dipartimento di Fisica Enrico Fermi, Università di Pisa, Largo Bruno Pontecorvo 3, I-56127 Pisa, Italy}
\affiliation{Istituto Nazionale di Ottica, Consiglio Nazionale delle Ricerche (CNR/INO), u.o.s. Adriano Gozzini, I-56127 Pisa, Italy}
\author{A. Macchi}
\affiliation{Dipartimento di Fisica Enrico Fermi, Università di Pisa, Largo Bruno Pontecorvo 3, I-56127 Pisa, Italy}
\affiliation{Istituto Nazionale di Ottica, Consiglio Nazionale delle Ricerche (CNR/INO), u.o.s. Adriano Gozzini, I-56127 Pisa, Italy}
\author{C. Riconda}
\affiliation{LULI, UPMC Université Paris 06: Sorbonne Universités, CNRS, Ecole Polytechnique, CEA, Université Paris-Saclay, F-75252 Paris Cedex 05, France}

\date{\today}

\begin{abstract} 
The Weibel instability driven by two symmetric counter-streaming relativistic electron plasmas, also referred to as current-filamentation instability, is studied in a constant and uniform external magnetic field aligned with the plasma flows. Both the linear and non linear stages of the instability are investigated using analytical modeling and Particle-In-Cell (PIC) simulations. While previous studies have already described the stabilizing effect of the magnetic field, we show here that the saturation stage is only weakly affected. The different mechanisms responsible for the saturation are discussed in detail in the relativistic cold fluid framework considering a single unstable mode. The application of an external field leads to a slighlty increase of the saturation level for large wavelengths, while it does not affect the small wavelengths.
Multi-mode and temperature effects are then investigated. While at large temperature the saturation level is independent of the external magnetic field, at small but finite temperature the competition between different modes in the presence of an external magnetic field leads to a saturation level lower with respect to the unmagnetized case.   
\end{abstract}


\maketitle

\section{Introduction}

The Weibel or current-filamentation instability has attracted extensive attention from both the astrophysics and laser-plasma communities. In astrophysics, it is believed to be the mechanism driving strong collisionless shocks in several astrophysical scenarios, from gamma-ray bursts and their afterglows, to the interaction of relativistic jets with the interstellar medium close to Active Galactic Nuclei, or in SuperNova Remnants, etc.~\cite{LarmorSaturation,Medvedev2005,WeibelShockGRB,Nishikawa2003,Spitkovsky2008}.
In the study of cosmic ray production, the Weibel instability is often quoted as the mechanism able to provide the intense magnetic field at the origin of charged particle scattering and their subsequent acceleration via the second order Fermi mechanism~\cite{FermiAcceleration,FermiAcceleration2}. 
Lately laser-plasma experiments have been able to identify the Weibel instability driven by two counter-straming high-energy flows~\cite{Fox2013,Park2015}. 
In this context, several numerical studies based on first principle simulations have been developed to study the physics of the Weibel instability~\cite{Fonseca2003} and of the Weibel-mediated shock in both astrophysics~\cite{Spitkovsky2008,KatoTakabe2008} and laser-plasma experiments~\cite{FiuzaPRL2012,Lobet}.  

Understanding the instability evolution in both its linear and nonlinear phases, and the prediction of the amplitude of the Weibel-generated magnetic fields are therefore of primary importance for a deeper insight into various astrophysical events, as well as for laser-plasma related studies. 

The Weibel instability has an electromagnetic nature and it can be triggered by small amplitude electromagnetic fluctuations. As charged particles get deflected by any fluctuation of the magnetic field perpendicular to their initial velocity, particles initially moving in opposite directions will concentrate in spatially separated current filaments hence amplifying the initial magnetic field perturbation (linear phase). As the self-generated magnetic field amplitude grows, the particle dynamics is strongly modified by the fields (non-linear phase), and various saturation mechanisms may set in. On a longer timescale, filaments of parallel currents tend to attract each other and merge, forming larger filaments (late merging phase).

 A Weibel unstable initial condition requires an anisotropy in the distribution function, that can be produced by a strong temperature anisotropy (the scenario originally envisioned by Weibel ~\cite{WeibelPaper}), or by counter-streaming flows (driving the then so-called current-filamentation instability). In both situations the instability transfers energy from the particles to the magnetic field and tends to isotropize the particle distribution function. 

This instability has been at the center of several recent works and different configurations have been investigated for unmagnetized plasmas ~\cite{Pegoraro1996,Califano1997,Califano1998}.
However background magnetic fields are present in various astrophysical environments where the instability is most likely to develop or have been proposed in laser-plasma experiments as a way to control and/or direct the high-energy plasma flows. Present studies Refs.~\cite{Wei_ChinesePL,StockemBcRel,Stockem2008} related to the magnetized scenarios, leave many open questions, in particular regarding the saturation mechanisms at play. 

In order to clarify some of these issues, this paper studies the case of a constant and uniform external magnetic field aligned with two counter-streaming relativistic electron flows in a neutralizing immobile ion background. Using both analytic modeling and Particle-In-Cell (PIC) simulations, we highlight the effect of the external magnetic field on the linear and nonlinear phases of the instability. In particular we show that, the well-known result~\cite{Wei_ChinesePL,StockemBcRel,Stockem2008} that the growth rate of the instability is reduced in the presence of a flow-aligned external magnetic field does not imply that the latter has a stabilizing effect in the nonlinear phase. Furthermore, we generalize previously proposed saturation mechanisms to account for the presence of an external magnetic field. We show that the magnetic field strength at saturation for a given wavenumber is weakly affected by the external magnetic field. 
Temperature and multi-mode effects are then investigated by seeding the instability from the electromagnetic fluctuations of a thermal plasma. At large temperatures we confirm that the saturation level is unaffected by the presence of the external magnetic field, while at lower (but finite) temperatures the external magnetic field mitigates the redistribution of the magnetic energy towards small wavenumbers, resulting in a saturation field somewhat smaller than without the external field.

The paper is structured as follows. Section~\ref{sec:SingleMode} considers the evolution of a single-mode with the cold relativistic plasma model. The linear phase is first considered analytically in Sec.~\ref{sec:LinearTheoryCold}. Theoretical predictions are compared with 1D3V PIC simulations in Sec.~\ref{sec:SimSetup1mode}. The various mechanisms responsible for the saturation of the instability are studied analytically in Sec.~\ref{sec:SaturationTheory} and tested against simulations in Sec.~\ref{sec:PICsaturation}.
Temperature and multi-mode effects are then investigated in Sec.~\ref{sec:MultiMode}. The linear phase is described within the framework of the relativistic warm fluid theory in Sec.~\ref{RelWarmFT} and theoretical predictions for the growth rate are compared to PIC simulations in Sec.~\ref{Sec:SimLinearTemp}. Sections~\ref{sec:SaturationMultiMode} and~\ref{Merging} discuss by means of PIC simulations the non-linear phase, saturation and late merging phase, respectively. Finally, Sec.~\ref{Conclusions} presents our conclusions.

\section{Cold relativistic plasma: single-mode theory and PIC simulations}\label{sec:SingleMode}

\subsection{Linear phase}\label{sec:LinearPhase}

\subsubsection{Relativistic cold fluid theory}\label{sec:LinearTheoryCold}

We will start first by an analytical description of the linear phase of the Weibel instability. It can be studied by taking ions at rest, providing a uniform neutralizing background for the two counter-streaming electron species (with respective densities $n_0/2$ and drift velocities ${\bf v}_0 = \pm v_0 {\bf {\bf\hat{z}}}$), modeled using a relativistic cold-fluid model~\cite{belmont2014}. 
We consider a uniform external magnetic field ${\bf B}_{0}=B_{0}{\bf\hat{z}}$ parallel to the initial electron plasma drift velocity. Linearizing the governing equations and considering all space-/time-dependent physical quantities $\phi(t,{\bf x}) \simeq \phi_0 \exp\big[-i(\omega t - {\bf k}_{\perp}\cdot{\bf x})\big]$, where $_\perp$ refers to the direction perpendicular to the flow, we obtain the dispersion relation for the purely transverse modes ~\cite{Wei_ChinesePL,StockemBcRel} 

\begin{equation} \label{eq:DispRelation}
\frac{\omega^{2}}{c^{2}}-{\bf k}_{\perp}^{2}-\frac{\omega_{p}^{2}}{c^{2}\gamma_{0}}\left(\frac{1}{\gamma_{0}^{2}}+\frac{{\bf k}_{\perp}^{2}v_{0}^{2}}{\omega^{2}-\Omega_{0}^{2}}\right)=0
\end{equation}
where $\omega$ and ${\bf k}_{\perp}$ are the frequency and wavevector of the considered modes, $\omega_p=\sqrt{4\pi e^{2}n_0/m_{e}}$ is the plasma frequency associated with the total density $n_0$, and $\Omega_{0}=-eB_{0}/(\gamma_{0} m_e c)$ is the cyclotron frequency of an electron in the external magnetic field ${\bf B}_0$. Gaussian-CGS units will be used throughout the paper.

The growth rate of the instability is found from the dispersion relation Eq.~\eqref{eq:DispRelation} where $\omega=i\Gamma$ with $\Gamma>0$. One then obtains
\begin{eqnarray} 
\nonumber \Gamma(k)&=&\frac{1}{\sqrt{2}}\left[\sqrt{\left(k^2c^2+\frac{\omega_p^2}{\gamma_0^3}-\Omega_0^2\right)^2+4\frac{\omega_p^2}{\gamma_0}k^2v_0^2} \right. \\ 
\label{eq:GrowthRate} &-&\left. \left(k^2c^2+\frac{\omega_p^2}{\gamma_0^3}+\Omega_0^2\right)\right]^{1/2}
\end{eqnarray}
where $k=|{\bf k}_{\perp}|$. 

In the limit of large wavenumbers $c^2k^2\gg\Omega_0^2+\omega_p^2/\gamma_0^3$, the growth rate takes the maximum and asymptotic value 
\begin{equation} \label{eq:GrowthRateAsymp}
\Gamma_{\rm max} = \sqrt{\frac{v_0^2}{c^2}\frac{\omega_p^2}{\gamma_{0}}-\Omega_{0}^{2}}\,.
\end{equation}
and we clearly see that $\Gamma$ is reduced by the external magnetic field $\Omega_0>0$.
Moreover, from Eq.~\eqref{eq:GrowthRate} we find that, in the presence of an external magnetic field, filaments with size larger than $\lambda_{\rm stab}=2\pi/k_{\rm stab}$ with
\begin{eqnarray}\label{eq:KstabCold}
k_{\rm stab}=\gamma_{0}^{-1}\left(\frac{v_{0}^{2}}{\Omega_0^{2}}-\gamma_{0}\frac{c^{2}}{\omega_{p}^{2}}\right)^{-1/2} 
\end{eqnarray}
cannot be created. Note that $k_{\rm stab}= \gamma_{0}^{-1}\left(r_L^2-d_e^2 \right)^{-1/2}$ with $r_L=v_0/\Omega_0$ the Larmor radius of an electron with velocity $v_0$ transverse to the external magnetic field and $d_e=c\sqrt{\gamma_0}/\omega_p$ the relativistic skin-depth. 
The growth rate indeed vanishes for $k \le k_{\rm stab}$ and only oscillatory solutions are admitted. 

From the above Eqs.~\eqref{eq:GrowthRateAsymp} and~\eqref{eq:KstabCold}, we easily find that there is a critical value of the external magnetic field above which the instability is quenched. The critical value is found by imposing $\Omega_0 =v_0\omega_p/(c\sqrt{\gamma_0})$ or $r_L=d_e$, for which $\Gamma_{\rm max}$ goes to zero and $k_{\rm{stab}}$ goes to infinity. The so-called critical magnetic field is given by
\begin{equation} \label{eq:Bcritical}
B_c=\sqrt{\gamma_0}\frac{v_0}{c}\frac{m_e\omega_p c}{e}\,.
\end{equation}
The condition $B_0 > B_c$ also corresponds to the case where the period of the electron gyration around $B_{0}$ is faster than the growth time of the instability computed in absence of the external magnetic field. For values of the magnetic field $0<B_0<B_c$, the formation of the filaments is slowed down.This can be explained considering that, once a particle is deflected in the direction perpendicular to the initial flow, toward the center of the filament, it starts gyrating around the external magnetic field. Similar considerations explain the stabilization of modes with large wavelengths, Eq.~\eqref{eq:KstabCold}. 

\begin{figure}[htb]
\centering 
\includegraphics[width=0.45\textwidth]{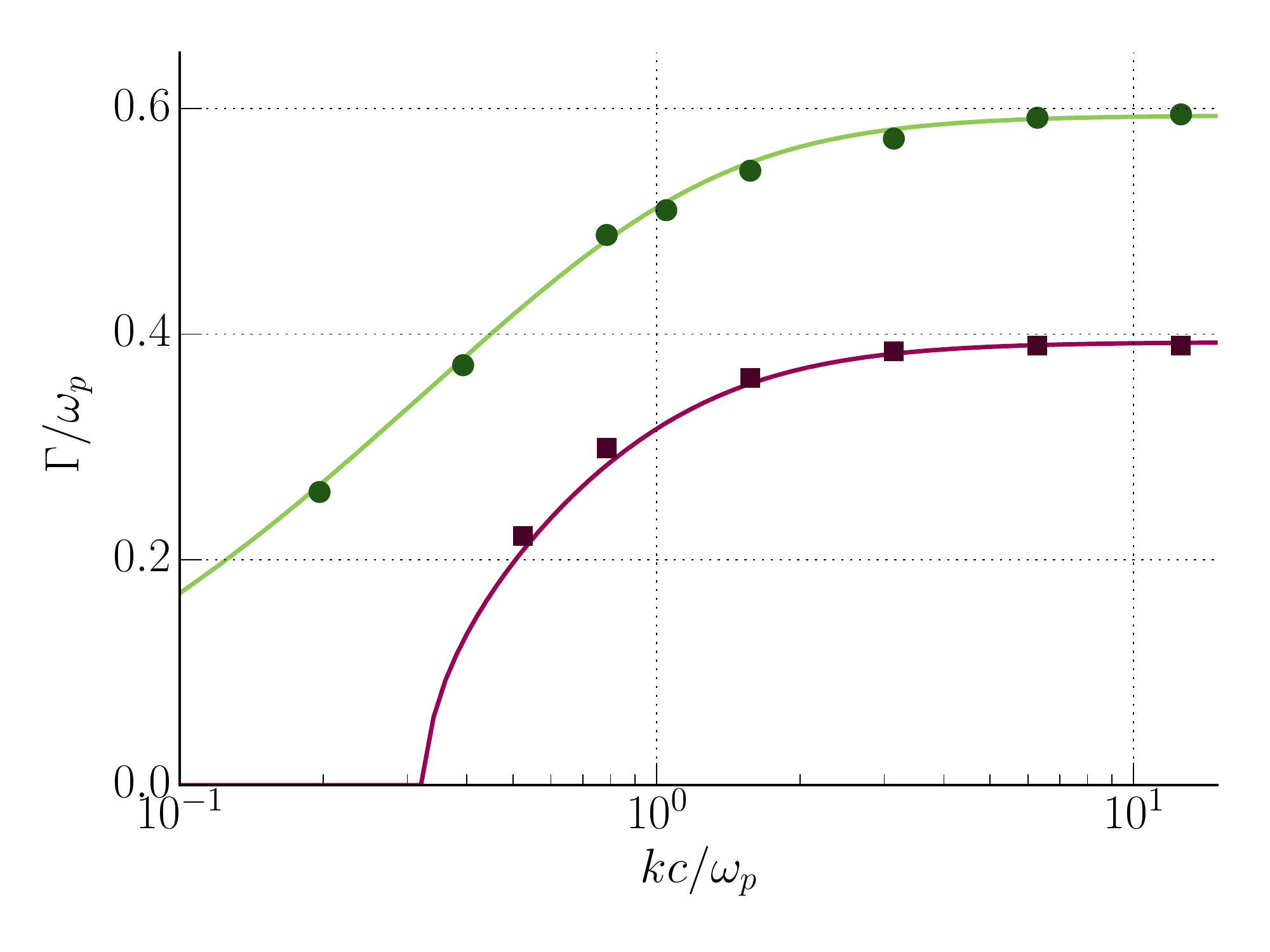}
\caption{\label{fig:GrowthRate} (color online) Growth rate of the instability as a function of the wavenumber for the cold plasma case (zero temperature). Analytical values for the unmagnetized (light green line) and magnetized $B_0=0.75\,B_c$ (dark purple line) cases are computed from Eq.~\eqref{eq:GrowthRate}. Circles (squares) correspond to the growth rate measured in 1D3V PIC simulations with a single-mode seeded perturbation and  $B_0=0$ ($B_0=0.75\,B_c$).}
\end{figure}

Figure~\ref{fig:GrowthRate} shows the growth rate $\Gamma(k)$ for electron flows with velocity $v_{z0}=\pm0.9c$ and external magnetic field $B_0=0.0$ (light green line) and $B_0=0.75\,B_c$ (dark purple line). For $B_0=0.75\,B_c$ (dark purple line), no unstable solutions are found for $k<k_{\rm stab}\simeq0.33\,\omega_p/c$, as predicted by Eq.~\eqref{eq:KstabCold}. We recall that without external magnetic field (light green line), the growth rate in the limit of small wavenumber $k^2c^2\ll \omega_p^2/\gamma_0^3$ increases linearly as $\Gamma(k)\simeq v_0\gamma_0 k$. 

In the rest of the paper we always consider this large value of the external magnetic field $B_0=0.75\,B_c$, in order to show that even if the growth rate is strongly reduced the saturation is not affected significantly.

Notice that for a given value of the external magnetic field, the maximum growth rate Eq.~\eqref{eq:GrowthRateAsymp} still depends on the electron drift velocity (or $\gamma_0$) and is reduced in the relativistic domain with increasing flow velocity. It takes its largest value for $\gamma_0=b_0^2+\sqrt{3+b_0^4}$, with $b_0=e B_0/(m_e c \omega_p)$. In the unmagnetized case $b_0=0$, this corresponds to $v_{0}\simeq0.82\,c$.

The linear theory also predicts that $E_z$, the inductive component of the electric field in the flow direction is phase-shifted with respect to the magnetic field (the maxima of $E_z$ being located at the nodes of $B_y$). It is proportional to $\Gamma(k)B_y/k$, and grows as fast as the magnetic field $B_y$. At this order there is no total density perturbation, and the electric field $E_x$ due to charge separation appears as a second order term.

\begin{figure}[htb]
\centering
\includegraphics[width=0.45\textwidth]{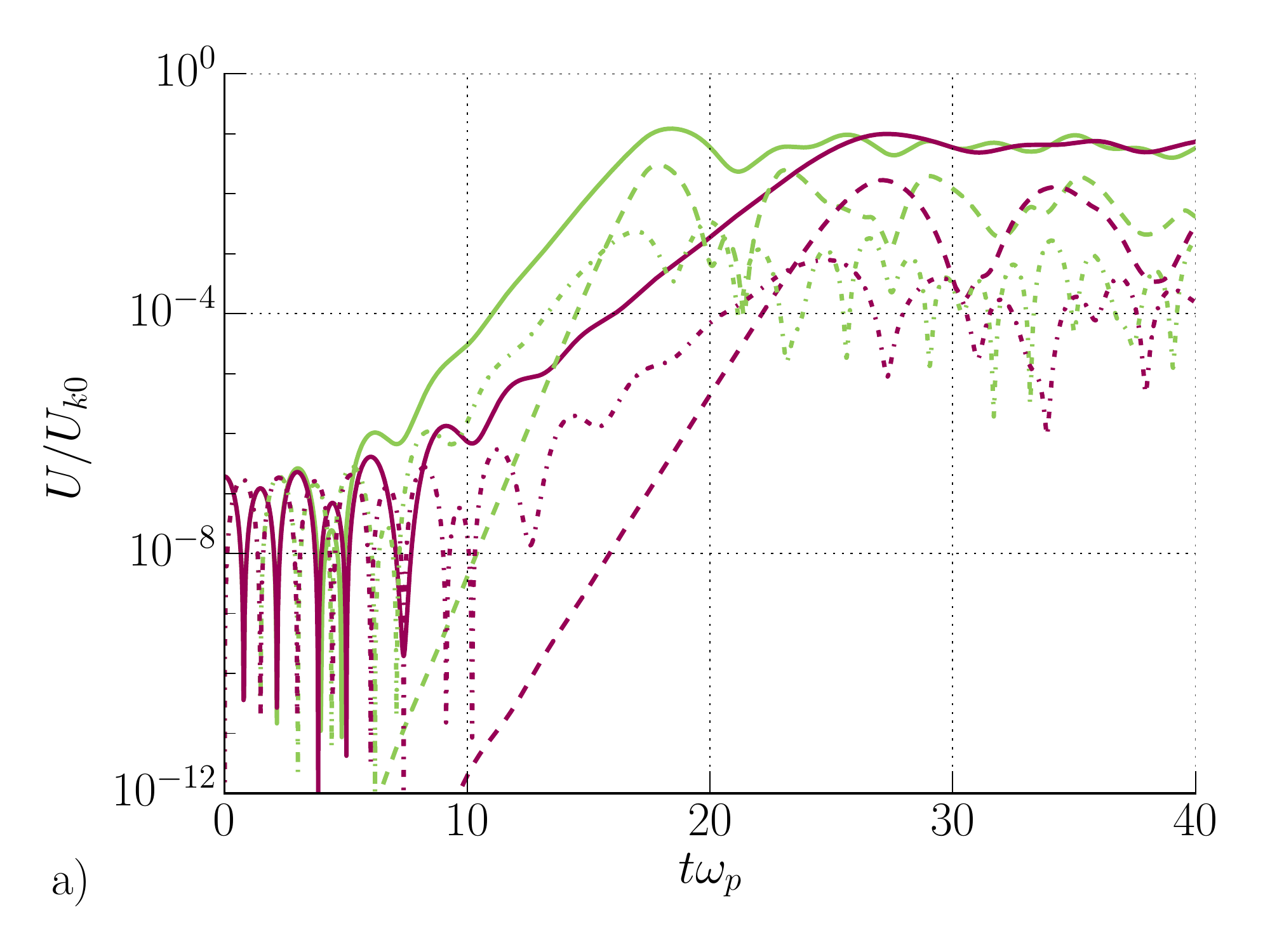}
\includegraphics[width=0.45\textwidth]{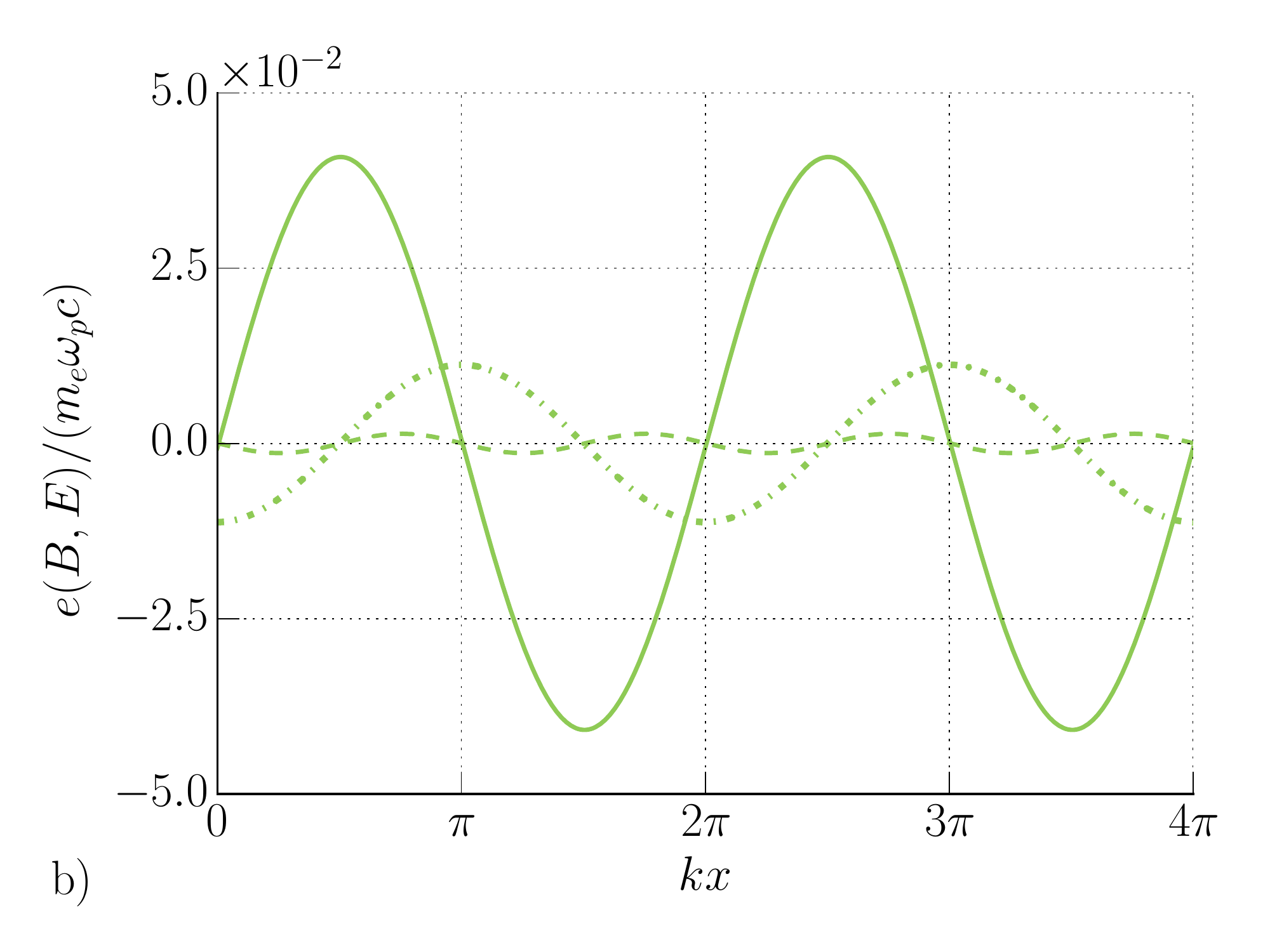}
\caption{\label{fig:EnergyGrowth} (color online) a) Evolution in time of the magnetic energy (plain line) and energies associated with the $E_z$ field (dash-dotted line) and $E_x$ field (dashed line) for the simulation with seeded mode $k=2.0\,\omega_p/c$. Light green lines refer to $B_0=0$ and dark purple lines to $B_0=0.75\,B_c$. All energies are normalized to the total initial flow energy $U_{k0}$. b) Spatial distribution of the magnetic field $B_y$ (plain line), electric field $E_z$ (dash-dotted line) and $E_x$ (dashed line) for the simulation without external magnetic field in the linear phase $t\simeq12\,\omega_{p}^{-1}$.}
\end{figure}

\subsubsection{Simulation set-up and comparison with theory}\label{sec:SimSetup1mode}

The analytical predictions of Sec.~\ref{sec:LinearTheoryCold} for the linear phase of the instability are confirmed by a series of 1D3V Particle-In-Cell (PIC) simulations. These are carried out in Cartesian geometry [${\bf x}=(x,y,z)$ and considering $\nabla = \partial_x {\bf{\bf\hat{x}}}$ in 1D3V] with the PIC code SMILEI~\cite{SmileiPaper}. 
The simulations also include the non-linear phase, discussed in the following Sec.~\ref{sec:PICsaturation}.
We consider two symmetric cold counter-streaming electron beams with initial drift velocities ${\bf v}_0=\pm v_0 {\bf\hat{z}}$ with $v_0=0.9\,c$ ( $\gamma_0\simeq2.3 $ mildly relativistic case).  Simulations with $\gamma_0=50$ (highly relativistic case) have also been performed, but the mildly relativistic case is representative of both situations, unless specified. 
The system has initially no net current. A population of immobile ions is taken into account in order to neutralize the total charge. 
In this 1D geometry, the Weibel instability amplifies the perturbations with wavevector ${\bf k}=k{\bf\hat{x}}$, magnetic field ${\bf B}=B_y{\bf\hat{y}}$ and inductive electric field ${\bf E}=E_z{\bf\hat{z}}$. 

In this Section, a single-mode is seeded as initial condition. This is done by imposing, at $t=0$, a magnetic field perturbation $B_{y0}(x)=\delta \sin(kx)$, with $\delta=0.001\,$ and $\lambda=2\pi/k$ the wavelength of the seeded mode. We consider wavenumbers in the range $0.2<kc/\omega_p<15$. 
The extension of the simulation box is $L_x=10\lambda$ and periodic boundary conditions are used. The resolution in space is $\Delta x=\lambda/200$ and in time is set to the $95\%$ of the CFL condition ($c\Delta t =0.95\,\Delta x$). The number of macro-particles-per-cell is $N_p=200$ for each species.

Figure~\ref{fig:EnergyGrowth}a shows the evolution in time of the energy in the magnetic field $B_y$, electric fields $E_z$ and $E_x$ for the simulation initialized with $k=2.0\,\omega_p/c$. Both unmagnetized ($B_0=0$, light green lines) and magnetized ($ B_0=0.75\,B_c$, dark purple lines) cases are presented. 
The phase of linear growth of the magnetic energy can be clearly identified in the interval $t=10-18\,\omega_p^{-1}$ $(t=15-28\,\omega_p^{-1}$) for the unmagnetized (magnetized) case. The values of the corresponding growth rates are reported in Fig.~\ref{fig:GrowthRate}. A very good agreement with the theory is obtained over the whole range of investigated $k$ values, for both the unmagnetized and magnetized cases. In particular, the growth rate of the instability is found to be reduced as $B_0$ is increased. Similar agreement has been found for $\gamma_0=50$ (not shown).

Figure~\ref{fig:EnergyGrowth}a also demonstrates the mainly magnetic nature of the Weibel instability despite $E_z$ growing as fast as the magnetic field $B_y$ [$E_z\sim\Gamma(k) B_y /k$].
During the linear phase, the space-charge electric field $E_x$ appears as a second order quantity. Indeed, it starts growing at a later time with respect to the magnetic component and it grows with twice the growth rate of the instability [Fig.~\ref{fig:EnergyGrowth}a]. The generation of this electrostatic field is a nonlinear effect, the onset of its growth corresponding to the formation of the current filaments resulting in a charge separation~\cite{Palodhi2009}. 
The $E_z$ component of the electric field is in counter-phase with the Weibel generated magnetic field, as shown in Fig.~\ref{fig:EnergyGrowth}b, and tends to reduce the current of the filaments slowing the particles down, as predicted from linear theory.

\subsection{Nonlinear phase and saturation}\label{sec:SaturationSingleMode}

\subsubsection{Theoretical considerations}\label{sec:SaturationTheory}

It has been well established in the literature ~\cite{LarmorSaturation, AchterbergNL} that, in the absence of an external magnetic field, two different mechanisms lead to the saturation of the Weibel instability.
At small wavenumber (large wavelength), saturation arises due to the so-called Larmor/Alfvén mechanism~\cite{LarmorSaturation}, while at large wavenumber (small wavelength) the so-called trapping mechanism is responsible for saturating the instability~\cite{AchterbergNL}. The generalization of these mechanisms to the magnetized plasma case is however not straightforward. In what follows, a single particle dynamics approach allows us to retrieve the saturation level predicted in the absence of an external magnetic field, while providing a better understanding of how these saturation mechanisms operate, and helps us to generalize these results to the magnetized case.

Let us consider the single particle dynamics in the fields developed during the linear stage of the instability. 
Despite the instability having a dominantly magnetic nature in its linear phase (see Sec.~\ref{sec:SingleMode}), we will consider the electron dynamics governed by the total magnetic field as well as by the inductive electric field $E_z$
\begin{eqnarray}
\label{eq:BlinTheory} {\bf B}(t,x)&=&B_{y0}\sin(kx)e^{\Gamma t}~{\bf\hat{y}} + B_0~{\bf\hat{z}}\,,\\
\label{eq:ElinTheory}{\bf E}(t,x)&=&-E_{z0}\,\frac{\Gamma}{ck}\cos(kx)e^{\Gamma t}~{\bf\hat{z}}\,,
\end{eqnarray} 
where $\Gamma=\Gamma(k)$ and $E_{z0} \sim \,B_{y0}$.
The equation of motion of an electron in the fields given by Eqs.~\eqref{eq:BlinTheory} and~\eqref{eq:ElinTheory} reads
\begin{align} 
\label{eq:B0SingleParticleEq_x} \frac{dx}{dt}&= \gamma_0v_0 \frac{\hat{p}_x(t)}{\gamma(t)} \\
\label{eq:B0SingleParticleEq_vx} \frac{d\hat{p}_x}{dt}&= -\hat{v}_z(t)\,\Omega_{y0}\sin(kx)e^{\Gamma t} +\hat{v}_y(t)\Omega_{0}  \\
\label{eq:B0SingleParticleEq_vy} \frac{d\hat{p}_y}{dt}&= -\hat{v}_x(t)\,\Omega_0\\
\label{eq:B0SingleParticleEq_vz} \frac{d\hat{p}_z}{dt}&= +\hat{v}_x(t)\,\Omega_{y0}\sin(kx)e^{\Gamma t} + \mathcal{E}_{z0}\,\cos(kx)e^{\Gamma t}\,,
\end{align}
where $\Omega_{y0}=-e B_{y0}/(\gamma_0 m_e c)$, $\mathcal{E}_{z0}=e E_{z0}\,\Gamma/(m_ec \gamma_0 v_0 k)$. In this section, momentum and velocities have been normalized such that $\hat{p}_i = p_i/(m_e\gamma_0 v_0)$ and $\hat{v}_i=v_i/v_0$ where the velocity $v_0$ is by definition positive $v_{z0}=v_z(t=0)=\pm v_0$. No general analytical solution can be given for this system of equations. Therefore, we first solve the system numerically, then we derive analytical solutions valid under some approximations. 
\begin{figure}\centering
\includegraphics[width=0.49\textwidth]{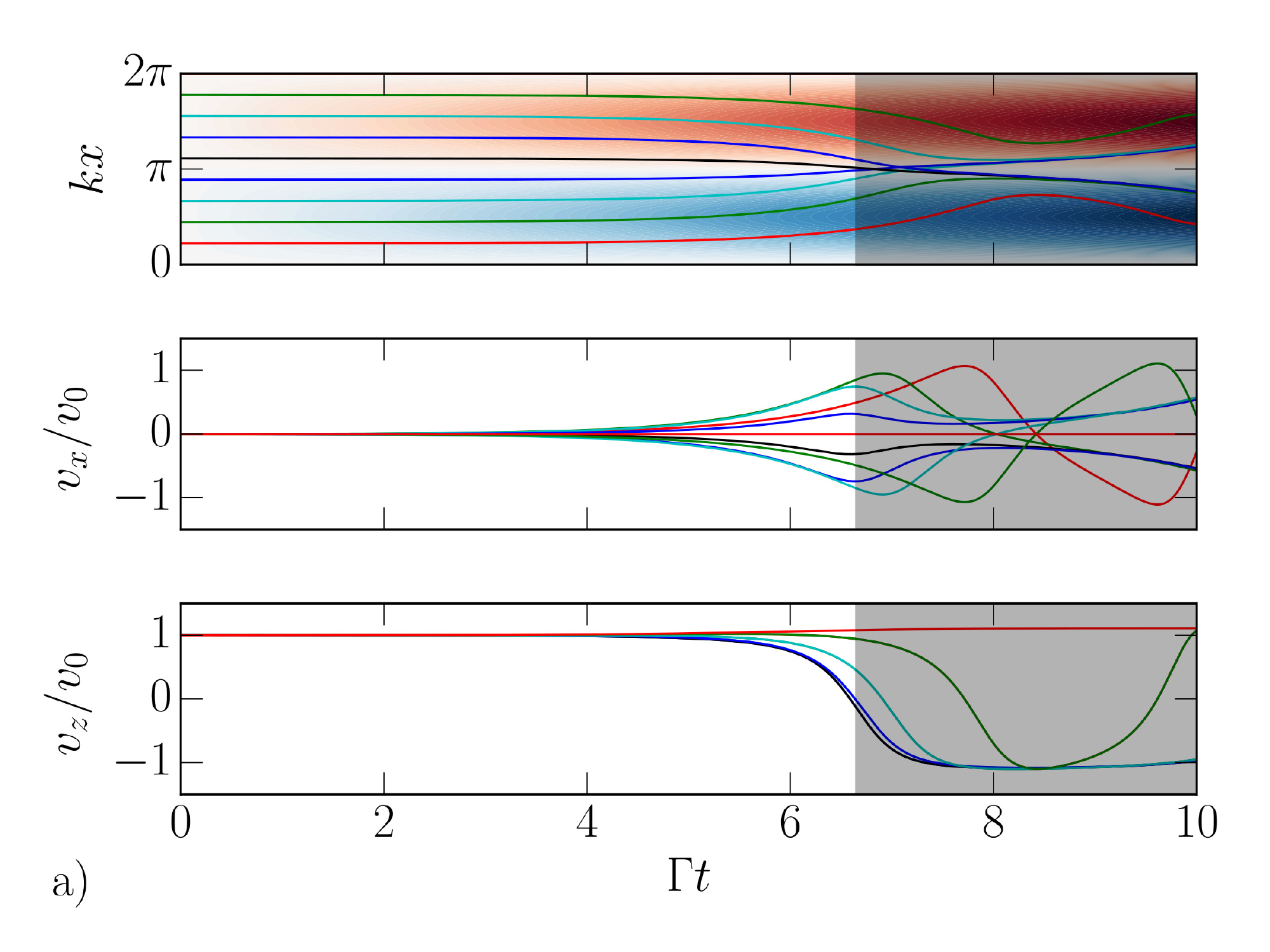}
\includegraphics[width=0.49\textwidth]{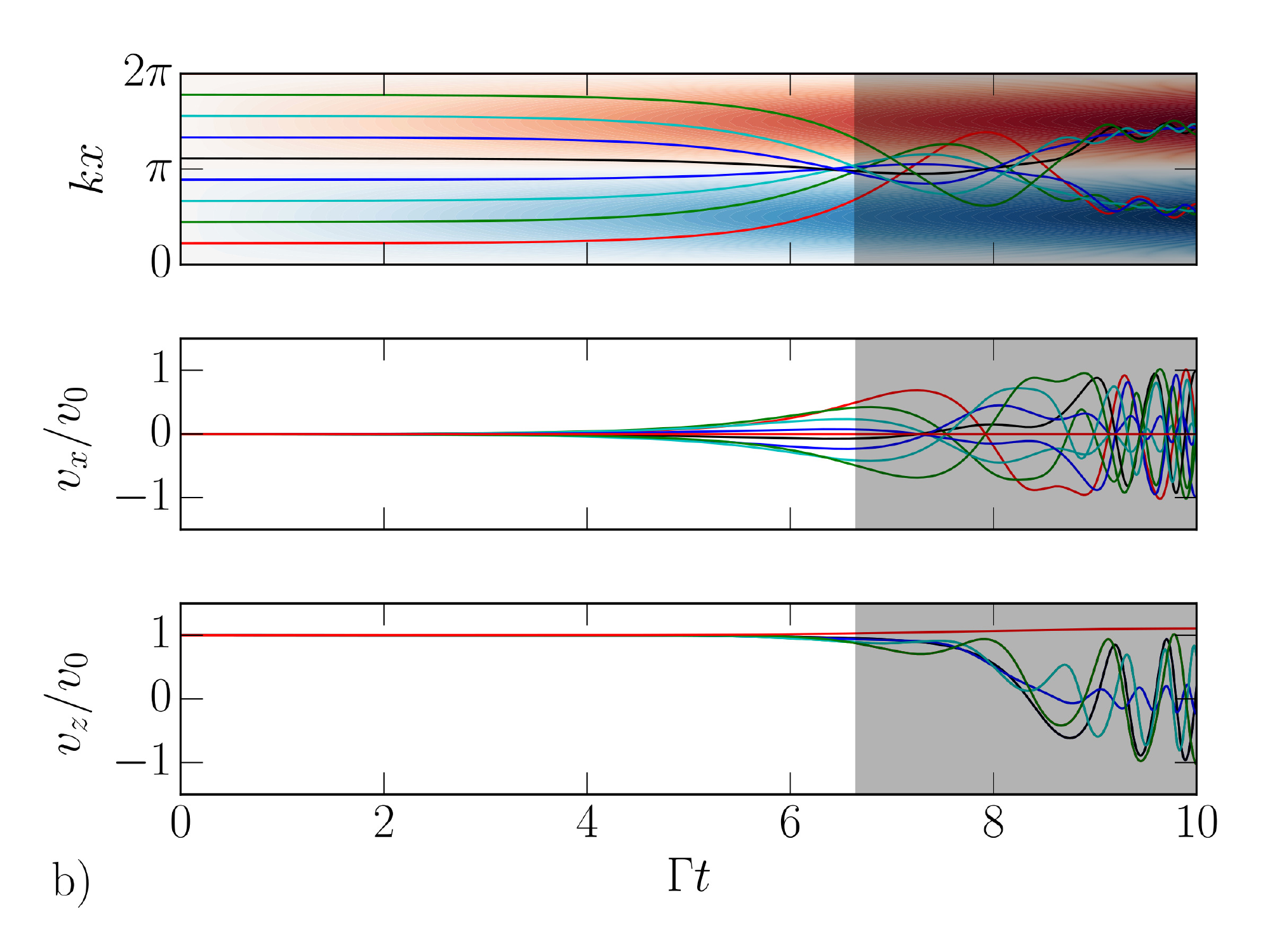}
\caption{\label{fig:NumPartTraj_g02.3}(color online) Typical trajectories of electrons with initial velocity ${\bf v}_0=+v_0 {\bf \hat{z}}$ ($\gamma_0=2.3$) in the electromagnetic fields developed during the linear stage of the instability: a) $k=0.35\,\omega_{p}/c$ (small-$k$), b) $k=2\,\omega_{p}/c$ (large-$k$). The trajectories are obtained numerically solving Eqs.~\eqref{eq:B0SingleParticleEq_x}-~\eqref{eq:B0SingleParticleEq_vz}. No external magnetic field is considered ($B_0=0$). In the top panel the red-blue colormap highlights the spatial distribution of the Weibel generated magnetic field. Blue (red) area corresponds to regions of positive (negative) $B_y$.}
\end{figure}

Typical electron trajectories obtained by numerically solving the system of Eqs.~\eqref{eq:B0SingleParticleEq_vx}-\eqref{eq:B0SingleParticleEq_vz} are given in Fig.~\ref{fig:NumPartTraj_g02.3} for the unmagnetized case $\gamma_0=2.3$, considering two values of the wavenumber $k=0.35\,\omega_{p}/c$ (Fig.~\ref{fig:NumPartTraj_g02.3}a) henceforth referred to as the small-$k$ case and $k=2\,\omega_{p}/c$ (Fig.~\ref{fig:NumPartTraj_g02.3}b) henceforth referred to as the large-$k$ case, corresponding to two different saturation mechanisms.

In both cases, the trajectories shown are those of electrons with an initially positive velocity $v_{z0}=v_0>0$, under the effect of the fields given by Eqs.~\eqref{eq:BlinTheory} and~\eqref{eq:ElinTheory}. These electrons will be mainly deflected toward the magnetic node in $k x=\pi$ and form what we will refer to as "the filament", the center of which being located at $k x=\pi$.

The numerical results are valid up to the saturation time $t=t_{\rm sat}$, at which $B_{\rm sat}=B_{y0}e^{\Gamma(k)t_{\rm sat}}$. In the Figure the dashed area corresponds to $t>t_{\rm sat}$ as deduced in the following section.  

The two different behaviors of the particle dynamics depending on their $k$ values are highlighted in Fig.~\ref{fig:NumPartTraj_g02.3}. 
In the small-$k$ case, electrons located at the center of the filament $k x \sim \pi$ see their longitudinal velocity $v_z$ decreased, even vanishing then changing sign. In contrast, in the large-$k$ case, all particles reach the center of the filament $k x=\pi$ with their velocity along the $z$-direction mainly unchanged $v_z \sim v_0$. 

The situation is totally symmetric if we consider particles with initial velocity $-v_0$, and the filaments form around $kx=0,2\pi$.\\

\paragraph{Saturation mechanism in the small-$k$ limit\\}

In the small-$k$ limit, saturation will be reached because particles inside the filament see their longitudinal velocity strongly reduced, hence decreasing the total current in the filament. In the absence of the external magnetic field, the saturation level can be recovered by equating the characteristic size of a filament $k^{-1}$ with the Larmor radius $r_L = v_0/\vert\Omega_{y,{\rm sat}}\vert$ of an electron with velocity $\pm v_0$ in the Weibel generated magnetic field. Similar estimates have already been derived in the literature considering that the saturation arises due to the Alfvén limitation of current~\cite{Alfven1939}. Indeed, as described in Appendix~\ref{AppendixAlfven}, there exists a maximum value of the current (Alfvén current) beyond which the longitudinal velocity $v_z$ of a particle initially at the border of the filament vanishes while crossing the center of the filament, and then reverses, due to the effect of the self-generated magnetic field. 
This estimate of the Alfvén limit however does not account either for the fact that, in the Weibel scenario, the magnetic fields are continuously and exponentially building up, nor for the effect of the resulting inductive electric field.
However by considering both these effects, we can show that we obtain the same saturation value that the (static) Alfvén limit.
From Eq.~\eqref{eq:ElinTheory}, we see that the inductive electric field is in counter-phase with the Weibel generated magnetic field, and has its maximum at the center of the filament. The dynamics of a particle initially located at the center of the filament $k x \sim \pi$ plays a central role in the saturation of the instability as shown in Fig.~\ref{fig:NumPartTraj_g02.3}. This position corresponds to a node of the magnetic field $B_y$, so that the particle dynamics will be marginally affected by the magnetic field.
It will be governed by the electric field $E_z$, leading to the reduced equation of motion, from Eq.~\eqref{eq:B0SingleParticleEq_vz}
\begin{eqnarray}\label{eq:motionEz}
\frac{d\hat{p}_z}{dt}&= -\mathcal{E}_{z0}\,e^{\Gamma t}\,.
\end{eqnarray}
Solving Eq.~\eqref{eq:motionEz} and taking for the saturation time the moment in which the longitudinal momentum vanishes $\hat{p}_z = 0$, allows one to derive  the strength of the magnetic field at saturation
\begin{eqnarray}\label{eq:Bsat_smallk}
B_{\rm sat}^{k\ll} = \gamma_0\,\frac{v_0}{c}\,\frac{c k}{\omega_{p}}\,\frac{m_e \omega_{p}c}{e}\,.
\end{eqnarray}
This is exactly the same value as obtained from the Alfvén current limitation or Larmor radius saturation. The single particle solution shows that taking into account only the temporal growth of the magnetic field $B_y$ and neglecting the induction field $E_z$, would overestimate the saturation level.
Indeed, Eq.~\eqref{eq:Bsat_smallk} is valid only if one considers both the fields $B_y$ and $E_z$, thus finally justifying the use of the static condition. 

The static Alfvén picture can be generalized to the case with external flow-aligned magnetic field. The calculations we performed in this configuration show that the saturation level increases with respect to the unmagnetized case. Considering a sinusoidal profile for the current and the magnetic field, in a 1D configuration, and calculating the field that corresponds to $\hat{p}_z=0$ for a particle moving toward the center, the predicted saturation value is
\begin{equation}\label{eq:Bsat_smallkB0}
B_{\rm sat}^{k\ll} = f(A) \gamma_0\frac{v_0}{c}\,\frac{c k}{\omega_{p}}\,\frac{m_e \omega_{p}c}{e} 
\end{equation}
with $f(A)=\left[\cos\left(\pi/2(1-A)\right)\right]^{-1}>1$ for $A<1$, and $f(A)=1$ for $A\ge 1$, with $A=v_0/\big| \Omega_0x_0\big|$ and $x_0=\lambda/4$ the particle initial position. Equation~\eqref{eq:Bsat_smallk} is recovered in the limit $A\gg1$. The detailed derivation is given in  Appendix~\ref{AppendixAlfven}.\\

\paragraph{Saturation mechanism in the large-$k$ limit\\}

Let us start with the unmagnetized case.
In the large-$k$ limit, the particle longitudinal velocity is mainly unchanged $v_z \sim v_0$ and thus the saturation follows from a different mechanism. 
Saturation is expected once all the particles have been injected inside the filament, i.e. once they have reached $k x=\pi$ (Fig.~\ref{fig:NumPartTraj_g02.3}). Thereafter no additional particles can be found to increase the current and contribute to the instability growth. The current of all the particles with velocity $v_0$ in one filament of extension $\sim \lambda/2$ remains much smaller that the Alfvén limit.
In the large-$k$ limit $\Gamma(k)/ck \ll 1$, the contribution of the longitudinal field $E_z$ can be neglected [see Eq.~\eqref{eq:ElinTheory}].
Indeed, numerically solving Eqs.~\eqref{eq:B0SingleParticleEq_x}-\eqref{eq:B0SingleParticleEq_vz} with or without $E_z$ (not shown) does not affect the particle trajectories.
Neglecting the effect of the electric field and considering $v_z \sim \pm v_0$, the system of Eqs.~\eqref{eq:B0SingleParticleEq_vx}-\eqref{eq:B0SingleParticleEq_vz} leads to an ordinary differential equation for the normalized particle position $\xi(t) = k x(t)$
\begin{eqnarray}\label{eq:xi}
\frac{d^2\xi}{d\tau^2} = -\alpha\,\sin(\xi)\,\exp(\tau)\,,
\end{eqnarray}
with $\tau = \Gamma t$ and $\alpha = {\rm sgn}\{v_{z0}\} \, v_0 k \Omega_{y0}/\Gamma^2$, with initial conditions $\xi(\tau=0)=$ $kx(t=0) = \xi_0$ and $d\xi/d\tau\vert_{\tau=0}=0$.

Considering a particle initially located at a maximum/minimum of the magnetic field $\xi_0^{\pm}=\pi \pm \pi/2$, leads to
\begin{eqnarray}
\xi^{\pm}(t) = \xi_0^{\pm} \mp \alpha\,\left[ \exp(\tau)-\tau-1\right].
\end{eqnarray}
The particle sees its velocity $v_x \propto e^{\tau}$ exponentially increasing with time, 
and depending on the sign of $v_{z0}$, the particle will head toward one or the other node of the magnetic field, hence spatially segregating particles with opposite velocities in well separated currents of opposite directions.
Taking the limit $\tau \gg 1$, one can extrapolate from this result the time $\tau^* \sim \ln\big(\pi/\vert 2\alpha\vert\big)$ at which the particle reaches the node of the magnetic field, and infer from this the corresponding magnetic field amplitude at saturation $B_{\rm sat}=B_{y0}\exp(\tau^*)$, leading to
\begin{eqnarray}\label{Bysat_largek}
B_{\rm sat}^{k\gg} = \frac{\pi}{2}\,\frac{\gamma_0 \Gamma^2_0}{v_0 k}\,\frac{m_e c}{e}\,.
\end{eqnarray}

Usually in the literature Ref.~\cite{AchterbergNL}, the magnetic field strength at saturation at large-$k$ is computed by equating the so-called bouncing frequency $\omega_b$ in the magnetic field at saturation with the growth rate of the instability~\cite{davidson1972}. Computing the bouncing frequency of an electron in the saturation field given by Eq.~\eqref{Bysat_largek} would indeed give
\begin{eqnarray}\label{eq:BouncingFreq}
\omega_b = \left(\frac{e v_0 k B_{\rm sat}^{k\gg}}{\gamma_0 m_e c}\right)^{1/2} \simeq \Gamma\,.
\end{eqnarray}
While Eq.~\eqref{Bysat_largek} leads a prediction similar to Eq.~\eqref{eq:BouncingFreq}, it highlights that saturation is obtained because all particles are injected and trapped into the filament.

The criterion on the bouncing frequency can be generalized in the presence of an external magnetic field. The bouncing frequency in this case becomes 
\begin{eqnarray}
\Omega_b = \sqrt{\omega_b^2+\Omega_0^2}
\end{eqnarray}
with $\omega_b$ defined in Eq.~\eqref{eq:BouncingFreq}. Considering that saturation is reached when the bouncing frequency equates the growth rate of the instability $\Gamma$, the expected saturation level would depend on the strength of the external magnetic field (see e.g. Ref.~\cite{Stockem2008}).  

To generalize the result of Eq.~\eqref{Bysat_largek} in the case of an external flow-aligned magnetic field, we calculate the saturation of the instability by considering that saturation occurs when all the electrons participate to the current filament. As a consequence, we can show that the saturation level of the instability is independent of the external magnetic field. 
We proceed as in the unmagnetized case: (i) we assume that the particle velocity is not drastically reduced at saturation $v_z(t) \sim \pm v_0$, and (ii) we neglect the effect of the longitudinal field $E_z$ on the particle motion, as $\Gamma(k)/k\ll 1$ in the large-$k$ limit. Both assumptions are found to be in good agreement with the numerical solution of Eqs.~\eqref{eq:B0SingleParticleEq_x}-\eqref{eq:B0SingleParticleEq_vz}, even in the presence of $E_z$.
One can write the equations of motion for a particle initially close to the maximum of the magnetic field, using $\sin(kx)\simeq 1$, in the form 
\begin{equation}\label{eq:dt2_vx_max}
\frac{d^2 v_x(t)}{dt^2}=v_0\Gamma \Omega_{y0}e^{\Gamma t}-\Omega^2_0 v_x\,.
\end{equation}
Looking for exponentially growing solution $v_x=v_{x0}e^{\Gamma t}$, as inferred from the unmagnetized case, the particle displacement $\delta x =x-x_0$ reads $\delta x=v_{x0}e^{\Gamma t}/\Gamma$. The saturation level is obtained for $\delta x(t_{sat})\simeq \lambda/4$, leading to
\begin{equation}\label{eq:TrappingB0}
B_{\rm sat}^{k\gg}= \frac{\pi}{2}\frac{\gamma_0 \left(\Gamma^2+\Omega^2_0\right)}{v_0 k}\frac{m_ec}{e}\,.
\end{equation}
In the limit $B_0=0$, we recover the result of Eq.~\eqref{Bysat_largek}. Moreover, Eq.~\eqref{eq:TrappingB0} predicts that the saturation level does not depend on the application of the external magnetic field for large-$k$. Indeed, the growth rate $\Gamma(k)$ decreases with the application of the external magnetic field [Eq.~\eqref{eq:GrowthRateAsymp}], but this variation is exactly compensated by the term $\Omega_0^2$ in Eq.~\eqref{eq:TrappingB0}, since the maximum value of the growth rate is $\Gamma^2 \sim \Gamma_0^2 - \Omega_0^2$, with $\Gamma_0$ the growth rate in the absence of external magnetic field. 
This is in contradiction with the estimate obtained considering the bouncing frequency but it is found to be confirmed by PIC simulations, as will be shown in Sec.~\ref{sec:PICsaturation}.

\subsubsection{Saturation phase in the PIC simulations} \label{sec:PICsaturation}

\begin{figure}
\centering
\includegraphics[width=0.45\textwidth]{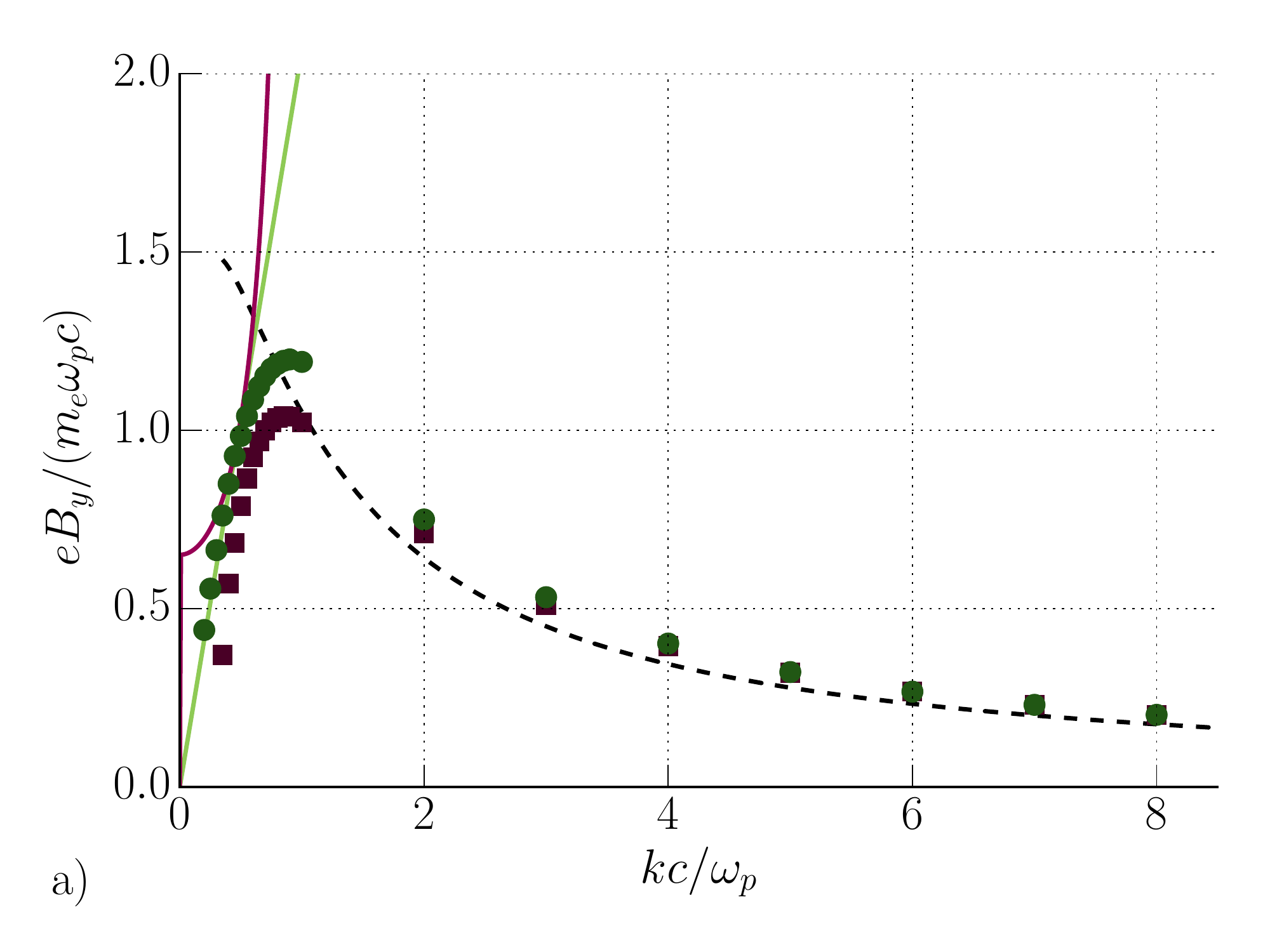}
\includegraphics[width=0.45\textwidth]{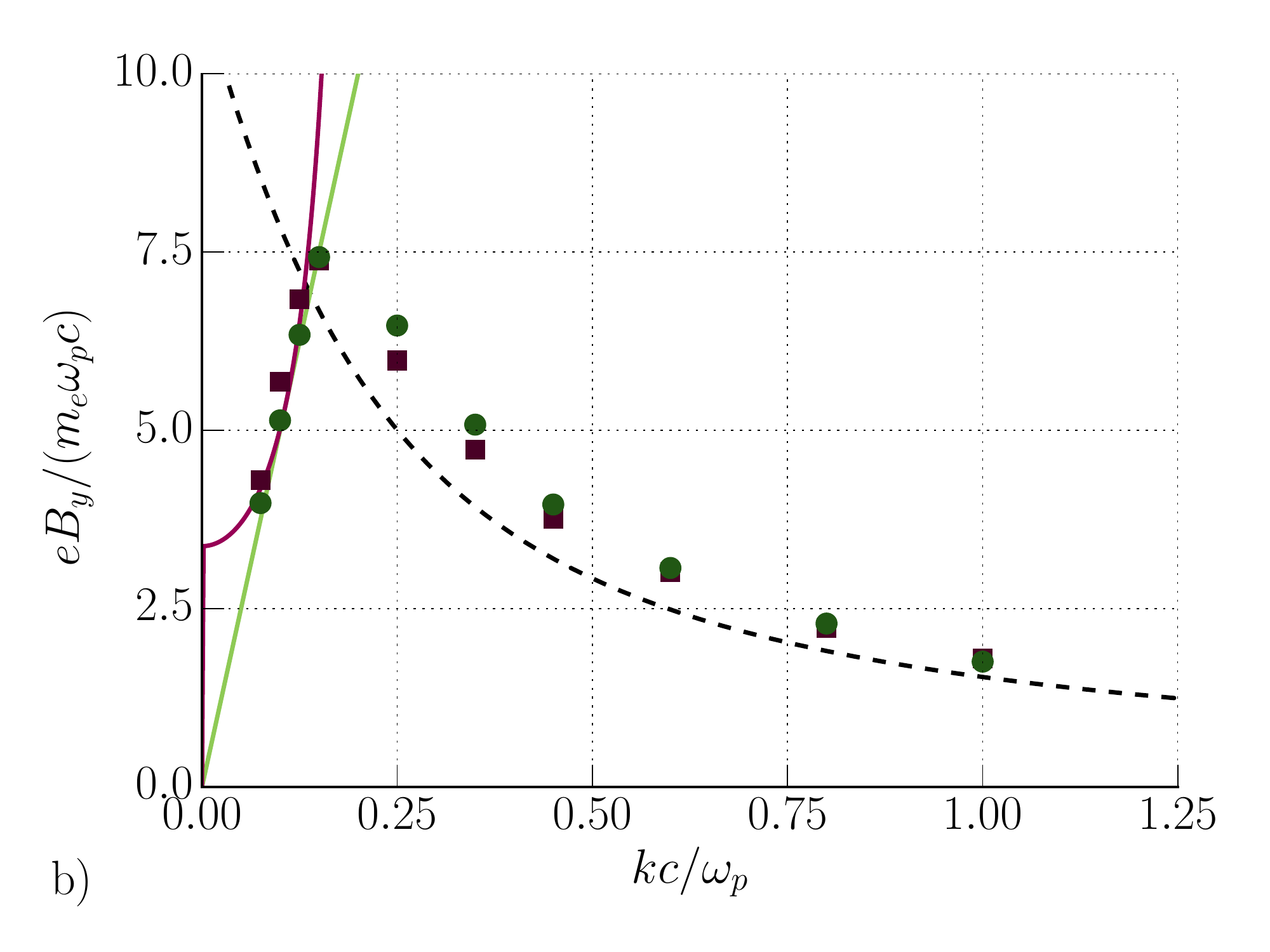}
\caption{\label{fig:SaturationMechanisms} (color online) Magnetic field strength at saturation. Values predicted by the "trapping mechanism" Eq.~\eqref{eq:TrappingB0} are shown as dashed line. Values predicted by the Alfvén limitation mechanism Eq.~\eqref{eq:Bsat_smallkB0} are shown as plain lines for the unmagnetized (light green) and for the magnetized (dark purple) cases. Circles (squares) are the values measured in PIC simulations seeded with a single-mode perturbation and $B_0=0$ ($B_0=0.75\,B_c$). Two initial flow velocities are considered: a)~$\gamma_0=2.3$ (mildly relativistic case), b)~$\gamma_0=50$ (ultra-relativistic case).}
\end{figure}

In this section we compare the theoretically predicted saturation level with the 1D3V PIC simulations presented in Sec.~\ref{sec:SimSetup1mode}.
The saturation levels are shown in Fig.~\ref{fig:SaturationMechanisms} as a function of the wavenumber, for two initial velocities corresponding to $\gamma_0\simeq2.3$ (mildly relativistic case) and $\gamma_0=50$ (ultra-relativistic case). We recall that these 1D3V simulations account for a single mode seeded at early time. In order to obtain the saturation level, we perform a Fourier spectrum of $B_y(x,t)$ and consider the maximum magnetic field of the given $k$ mode.  

Each unstable mode saturates because of the mechanism that predicts the lower saturation value. The maximum magnetic field is found at the intersection between the curves corresponding to the Alfvén limit Eq.~\eqref{eq:Bsat_smallkB0} and the trapping mechanism Eq.~\eqref{eq:TrappingB0}, respectively $k^*\simeq0.63\,\omega_p/c$ and $k^*_{B_0}\simeq 0.60\,\omega_p/c$ for the mildly relativistic case and for the ultra-relativistic case $k^*\simeq0.14\,\omega_p/c$ and $k^*_{B_0}\simeq0.13\,\omega_p/c$. 
It is clear that the Alfvén limit cannot be the dominant saturation mechanism for large wavenumbers. 
This can be easily understood as the magnetic energy (increasing with $k$) would exceed the total kinetic energy of the beams. The saturation would appear for lower values due to the trapping mechanism. Nevertheless for the modes with small $k$, the Alfvén mechanism is responsible for the saturation of the instability. 

Figure~\ref{fig:SaturationMechanisms} reports the measured saturation level for different unstable modes, for the two initial velocities in unmagnetized plasma and with $B_0=0.75\,B_c$. 
The trapping saturation mechanism is the dominant one for $k>k^*$. In this regime the theoretical predictions of Eq.~\eqref{eq:TrappingB0} show a very good agreement with the simulations, confirming the independence of the saturation level from the external magnetic field.  
For wavenumbers $k<k^*$ and $B_0=0$, the Alfvén limit accurately reproduces the data. In the magnetized case two different behaviors are observed for highly relativistic ($\gamma_0=50$) and mildly relativistic ($\gamma_0\simeq2.3$) flows. 
In the first case the saturation level is slightly increased, as predicted by the generalized Alfvén limit in a magnetized plasma [Eq.~\eqref{eq:Bsat_smallkB0}]. On the contrary in the mildly relativistic case Fig.~\ref{fig:SaturationMechanisms}a, the saturation level decreases with the application of $B_0$.

The discrepancy between Eq.~\eqref{eq:Bsat_smallkB0} and the numerical simulations in the mildly relativistic case is due to the fact that the single-mode analysis does not hold anymore. 
With the application of the external magnetic field, the growth rate is decreased and increase of the time required to reach saturation is increased. This results in the harmonics of the initial $k$ becoming important before the considered (seeded) mode reaches its saturation. In particular, we observe the growth of the third harmonic with a growth rate close to three times the one of the seeded mode $\sim 3\Gamma(k)$. This prevents the seeded mode reaching its own (independent) saturation level. This effect is strongly reduced in the ultra-relativistic limit where in the simulations a much weaker signal for the third harmonic is observed.

Analysis of the mildly relativistic simulations confirms that the saturation via the Alfvén limit is not reached: the velocity along the flow direction does not vanish. 
The total energy that is expected to be transferred to the magnetic field is instead distributed in the two modes, the seeded one with wavenumber $k$ and the harmonic at $3k$. 

In presence of harmonics, the single mode saturation criterion Eq.~\eqref{eq:Bsat_smallkB0} cannot be applied. 
However we can consider that saturation is associated to a redistribution of kinetic energy into magnetic field energy and the overall level of conversion into one mode and its harmonics has to be roughly the same as in the single mode case. It is then useful to calculate the ratio of the magnetic energy density over the kinetic energy. Indeed the Alfvén limit Eq.~\eqref{eq:Bsat_smallk} can also be interpreted as an energy equipartition relation for the most unstable mode ($kc/\omega_p\sim1$), the equipartition condition being defined as 
\begin{equation}
\frac{B_{\rm sat}^2/8\pi}{n_0(\gamma_0-1)m_ec^2}=\frac{1}{2}
\end{equation}
In reality the saturation level of Fig.~\ref{fig:SaturationMechanisms} gives an energy ratio for the most unstable $k$ saturating via the Alfvén mechanism, Eq.~\eqref{eq:Bsat_smallkB0}, smaller than $15\%$ for the mildly relativistic case and $10\%$ for the ultra-relativistic one, roughly independent from the external magnetic field. Similar levels of equipartition were already observed in simulations Refs.~\cite{Califano1998,LarmorSaturation}.
The predicted equipartition level in the mildly relativistic case for $k=0.35\,\omega_p/c$ (representative of the small-$k$ limit) is $\sim2\%$ as calculated with $B_{\rm sat}$ from Eq.~\eqref{eq:Bsat_smallkB0}. This is much larger than the value one would obtain considering the single mode saturation (value of $B_{\rm sat}$ as in Fig.~\ref{fig:SaturationMechanisms}a) but it is comparable ($1.8\%$) if the contributions of the two modes are considered. 

\begin{figure}
\centering 
\includegraphics[width=0.45\textwidth]{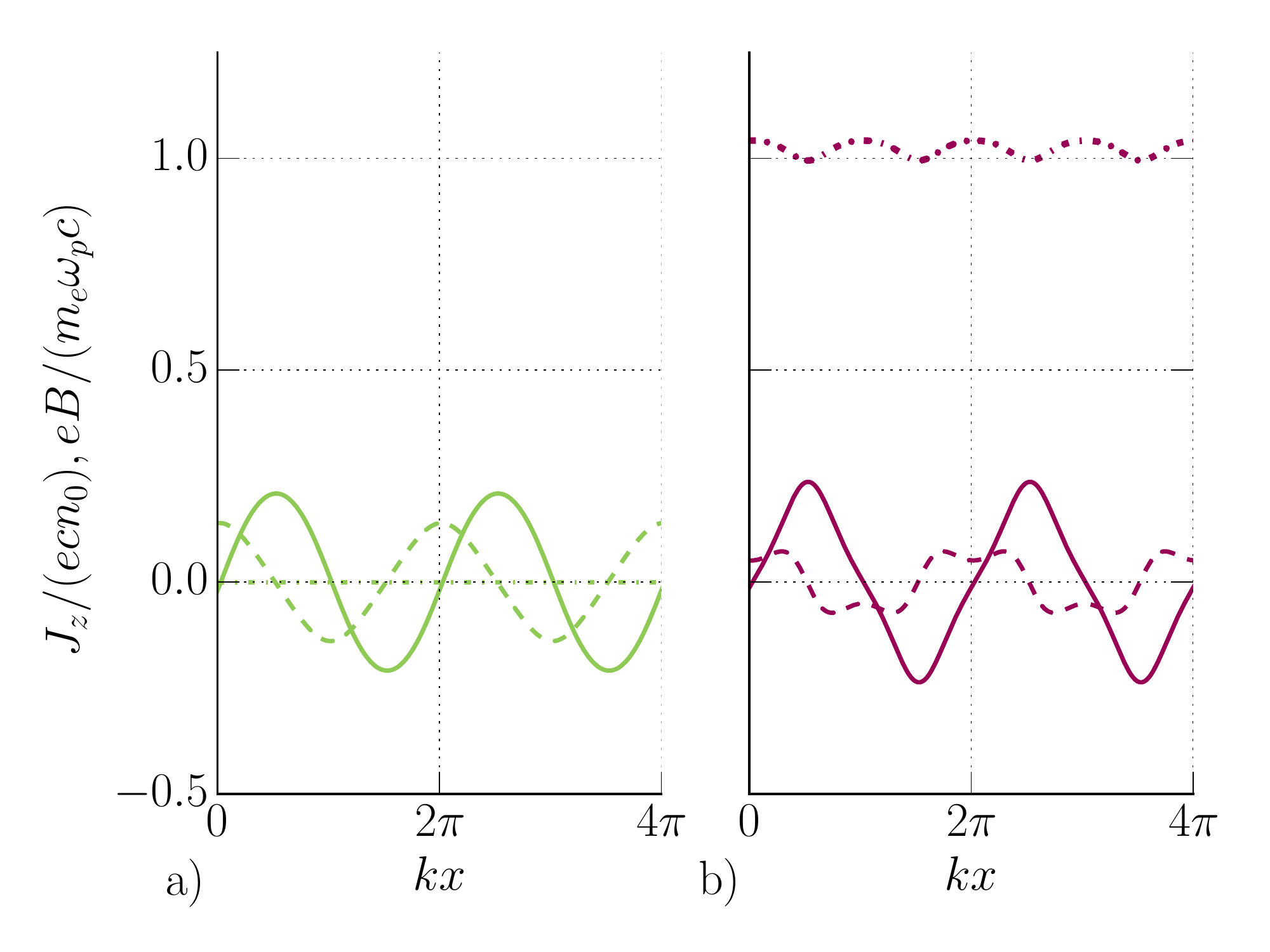}
\caption{ \label{fig:CurrentFilaments}(color online) Spatial distribution of the total current $J_z$ (dashed lines) in the initial direction of the flows, Weibel-generated magnetic field $B_y$ (plain line) and flow-aligned magnetic field $B_z$ (dash-dotted lines) for the mildly relativistic ($\gamma_0=2.3$) simulation in the small-$k$ limit $k=0.35\,\omega_p/c$ in the linear phase. a) unmagnetized case $B_0=0$ at $t\simeq 12.5\,\omega_p^{-1}$, b)~magnetized case $B_0=0.75\,B_c$ at $t\simeq\,26.5\omega_p^{-1}$.  }
\end{figure}

Since the harmonic is weaker in the ultra-relativistic case, the agreement with the theoretical curve is significantly improved, Fig.~\ref{fig:SaturationMechanisms}b.
This confirms that Eq.~\eqref{eq:Bsat_smallkB0} is only valid for single mode.
In the presence of higher harmonics the current filament profile evolves from a sinusoidal shape to a double peaked structure, see Fig.~\ref{fig:CurrentFilaments}b, where current filaments are formed of two consecutive maxima or minima. The electron density has the same profile as the current $J_z$, meaning that the particles are concentrated in the two spikes at the edge of the filament, and the hypothesis of sinusoidal profile used to derive Eq.~\eqref{eq:Bsat_smallkB0} breaks down.
The competition between different modes will be addressed in more detail in Sec.~\ref{sec:MultiMode}.

\begin{figure}
\includegraphics[width=0.6\textwidth]{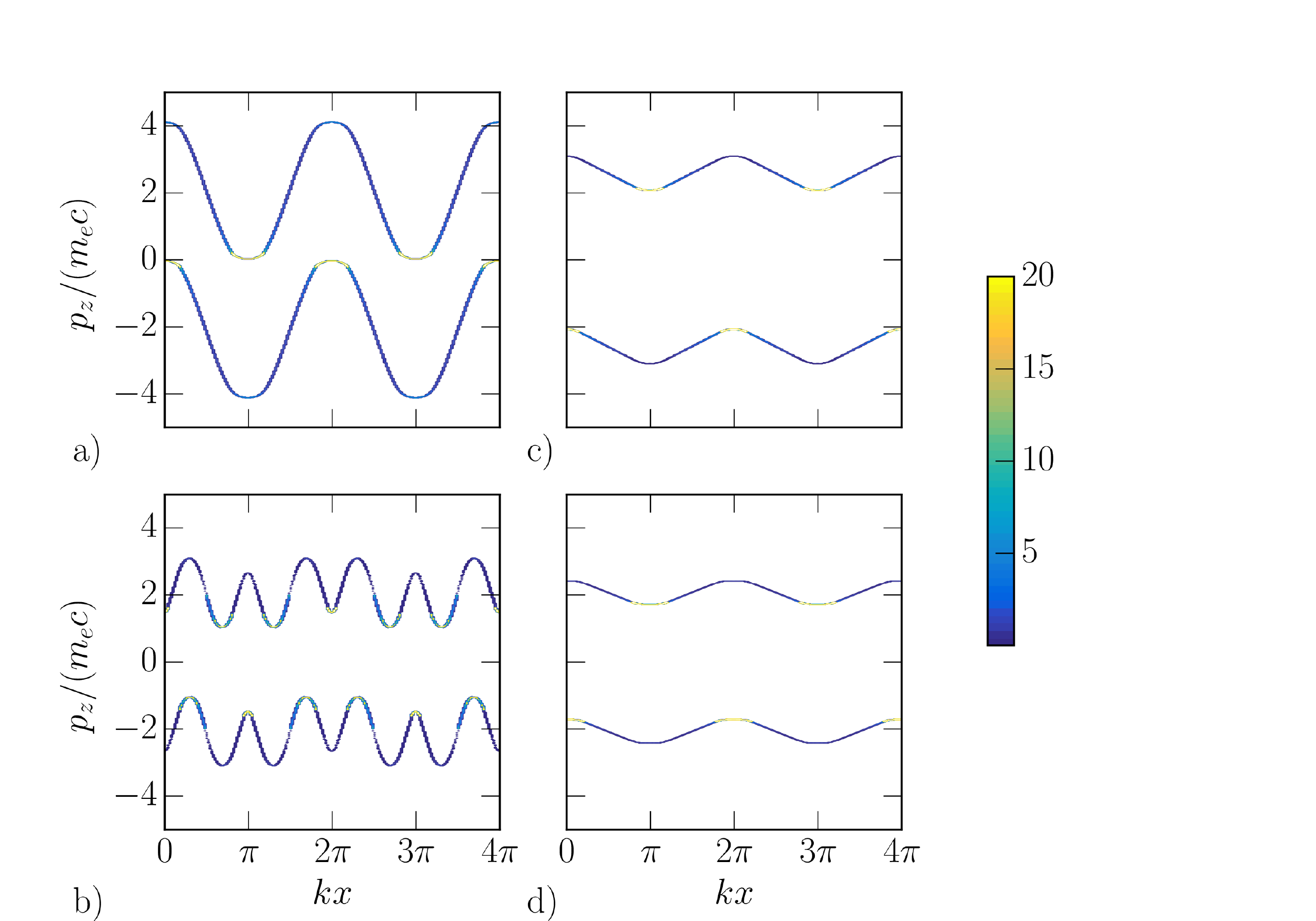}
\caption{\label{fig:SaturationPhaseSpace}(color online) ($x,p_{z}$)-phase space distribution at the saturation for the simulations initialized with a single mode in the mildly relativistic case $\gamma_0=2.3$. In the small-$k$ limit, $k=0.35\,\omega_p/c$: a) $B_0=0$ , b)  $B_0=0.75\,B_c$. In the large-$k$ limit, $k=2.0\,\omega_p/c$: c) $B_0=0$ , d)  $B_0=0.75\,B_c$. }
\end{figure}

The signature of the two different saturation mechanisms can be clearly observed in the PIC simulations.
Figure~\ref{fig:SaturationPhaseSpace} shows the phase space $x-p_z$ for the simulations with $\gamma_0=2.3$, for a small-$k$ mode ($k=0.35\,\omega_p/c$) and for a large one ($k=2\,\omega_p/c$) with and without external magnetic field, at the time corresponding to their own saturation. We chose two modes that saturate at the same value of $B_{y}$ but for the two different mechanisms, the Alfvén limit for small-$k$ and the trapping mechanism for the large-$k$. 
With large wavenumber, Figs.~\ref{fig:SaturationPhaseSpace}c,~\ref{fig:SaturationPhaseSpace}d, the flow kinetic energy associated with the motion along the ${\bf\hat{z}}$-direction is still large at saturation, and the value of $p_z$ is close to the initial one $p_{z}(t=0)\simeq\pm 2.1m_{e}c$, typical of the trapping mechanism. 
In the case of small-$k$ and $B_0=0$, the particles responsible for the saturation lye in the region $p_z \simeq0$, Fig.~\ref{fig:SaturationPhaseSpace}a, as expected from the Alfvén limit. As already discussed, in the mildly relativistic case, adding the external magnetic field, the harmonics of the seeded mode set in before the mode saturates. These harmonics are clearly seen in Fig.~\ref{fig:SaturationPhaseSpace}b.
The gain of momentum along the $z$-direction up to twice the initial value, observed in Fig.~\ref{fig:SaturationPhaseSpace} for all the simulations, is associated with the particles trapped in the region occupied by the filaments flowing in the opposite direction, as previously observed in Ref.~\cite{DAngeloMarta} in the case of counter-propagating electron-positron plasmas. 

\section{Temperature and multi-mode effects}\label{sec:MultiMode}

The introduction of an initial temperature has two effects. On the one hand it affects the single mode growth rate, on the other hand it allows for the growth of a broad spectrum of magnetic perturbations from the intrinsic electromagnetic fluctuations of a thermal plasma. The modification of the growth rate can be studied for a single mode and compared with PIC simulations in the linear phase when all the modes grow independently. However, saturation of the instability most often involves multi-mode evolution. We present the studies of the linear and non-linear phases in the following sections. 

\subsection{Linear Phase}\label{sec:LinearPhaseMultiMode}

\subsubsection{Relativistic warm fluid theory}\label{RelWarmFT}

The effect of the temperature in the linear phase of the instability has been investigated in the magnetized non relativistic regime in Refs.~\cite{Tautz_NRWarmGRB0, Bornatici1970} and in the relativistic one for the unmagnetized case in Ref.~\cite{BretPRE2010}. 
For the sake of analytical tractability, we use a relativistic fluid approach including the pressure of the relativistic plasma flows
\begin{eqnarray}
\partial_t n_{\pm} + \nabla\!\cdot\!(n_{\pm}\bf{v}_{\pm})\!\!\!&=&\!\!\!0 \\
h(\mu_{\pm})\!\left[ \partial_t \bf{p}_{\pm} +(\bf{v}_{\pm}\!\cdot\!\nabla)\bf{p}_{\pm}\right]\!\!\!&=&\!\!\!-e\!\!\left[{\bf E}\!+\!\frac{{\bf v}_{\pm}}{c}\!\times\!\bf{B}\right]\!-\!\frac{\nabla P_{\pm}}{n_{\pm}}\quad\,\,\,
\end{eqnarray}
where $\pm$ denotes the electron plasmas with initial velocity ${\bf v}_0=\pm v_0\hat{\bf z}$. In the following we consider symmetric counter-propagating beams.
The normalized enthalpy $h(\mu)$ depends on $\mu=m_ec^2/T$, with $T$ the rest frame plasma temperature. 
The closure of the fluid equations is done assuming an ideal gas $P=nT$. For small-$k$ ($k\sqrt{T/m_e}\ll\Gamma $) we expect the adiabatic ($T\propto n^{\eta-1}$) description to be valid, with $\eta$ the adiabatic index.
As a reference we also consider the isothermal closure ($T\,\rm constant$), as for larger $k$ the adiabatic closure might not apply.
Note that the enthalpy, that is often neglected, gives an important correction for $T\gtrsim m_ec^2$. 

In order to properly describe a plasma with arbitrary flow velocity and temperature, we use a Maxwell-Jüttner distribution function defined as~\cite{WrightPRA1975} 
\begin{equation} \label{eq:Maxwell-Juttner}
f_{\pm}({\bf p})=\frac{n_0\mu}{4\pi\gamma_0 K_2(\mu)} \exp\!\!\left[\!-\mu\gamma_0\!\left(\!\!\sqrt{1+\frac{{\bf{p}}^2}{m_e^2c^2}}\mp\frac{v_{0}p_z}{m_ec^2}\!\!\right)\!\right]
\end{equation}
with $K_n$ the modified Bessel function of the second kind. 
From this we obtain the normalized enthalpy of each beam $h(\mu)=K_3(\mu)/K_2(\mu)$ and adiabatic index $\eta(\mu)=1+1/\left[\mu h(\mu)-\mu-1\right]$~\cite{Melzani}. In the limit $T\ll m_e c^2$ (correspondingly $\mu \gg 1$) $h\simeq1$ and $\eta \simeq 5/3$, while for $T\gg m_e c^2$ ($\mu \ll 1$) $h\simeq4T/\left(m_ec^2\right)$ and $ \eta \simeq 4/3$. 

Assuming the enthalpy to depend on the initial rest frame temperature only, one proceeds as in Sec.~\ref{sec:LinearPhase} and obtains the dispersion relation for the purely transverse (Weibel) modes  
\begin{equation} \label{eq:DispRelationTemp}
\frac{\omega^{2}}{c^{2}}-k^{2}-\frac{\overline{\omega}_{p}^{2}}{c^{2}\gamma_{0}}\left(\frac{1}{\gamma_{0}^{2}}+\frac{v_{0}^{2}k^{2}}{\omega^{2}-\overline{\Omega}^{2}(k)}\right)=0\, ,
\end{equation}
where $\overline{\omega}_{p}^{2}=\omega_{p}^{2}/h(\mu)$ ,  $\overline{\Omega}^{2}(k) =\overline{\Omega}_0^{2}+\gamma_0^{-1}\eta(\mu)\overline{v}_{th}^2k^2$
with $\overline{\Omega}_{0}^{2}=\Omega_{0}^{2}/h^2(\mu)$ and 
$\overline{v}_{th}=\left[\mu h(\mu)\right]^{-1/2}$. 
From Eq.~\eqref{eq:DispRelationTemp}, we derive the growth rate of the instability 
\begin{eqnarray}
\nonumber \Gamma(k)&=&\frac{1}{\sqrt{2}}\left[\sqrt{\left(k^2c^2+\frac{\overline{\omega}_p^2}{\gamma_0^3} -\overline{\Omega}^2 (k)\right)^2 +4\frac{\overline{\omega}_p^2}{\gamma_0}k^2v_0^2}\right. \\
\label{eq:GrowthRateTemp} &-&\left.\left(k^2c^2+\frac{\overline{\omega}_p^2}{\gamma_0^3} + \overline{\Omega}^2(k)\right)\right]^{1/2}\, .
\end{eqnarray}
In the limit $T=0$, we recover the prediction of the  cold fluid theory, Eq~\eqref{eq:GrowthRate}. From Eqs.~\eqref{eq:DispRelationTemp} and~\eqref{eq:GrowthRateTemp}, we can deduce the range of unstable wavenumbers. The main effect of the temperature is to strongly reduce the instability growth rate at large $k$. The instability is completely quenched for wavenumbers larger than 
\begin{widetext}
\begin{equation}
c^2k^2_{\rm cut-off}= \frac{\gamma_0}{2 \eta\overline{v}_{th}^2}\left[\frac{\overline{\omega}_p^2v_0^2}{\gamma_0c^2}-\frac{\overline{\omega}_p^2
\eta\overline{v}_{th}^2}{\gamma_0^4c^2}-\overline{\Omega}^2_0 +\sqrt{ \left( \frac{\overline{\omega}_p^2v_0^2}{\gamma_0c^2}-\frac{\overline{\omega}_p^2 \eta\overline{v}_{th}^2}{\gamma_0^4c^2}-\overline{\Omega}^2_0 \right)^2-4\frac{\overline{\Omega}_0^2\overline{\omega}_p^2 \eta\overline{v}_{th}^2}{\gamma_0^4c^2}}\right]\,.
\end{equation}
\end{widetext}
Indeed, thermal motion of the particles in the direction transverse to the flow prevents their confinement in the filaments.
Figure~\ref{fig:GrowthRateTemperature} shows the growth rate of the instability as a function of the wavenumber for the two possible closures of the fluid equations. 
The range of modes amplified by the instability is clearly dependent on the temperature: the higher the temperature, the smaller the value of $k_{\rm cut-off}$.
Qualitatively, the difference between unmagnetized/magnetized systems can be explained considering that the growth rate of the instability decreases with the introduction of $B_0$. In order to allow the instability to grow, a particle should remain in the region where the filament forms for a time of the order of $\Gamma^{-1}$. The larger the external magnetic field, the longer the required interval of time. Hence with equal temperatures the small filaments are less likely to form in the magnetized case, and the value of $k_{\rm cut-off}$ decreases. 

\subsubsection{PIC simulation set-up and comparison with linear theory}\label{Sec:SimLinearTemp}

In order to investigate the temperature effects and the interplay between the various growing modes, we present a series of 1D3V simulations with, at initial time, a broad spectrum of modes  seeded exploiting the intrinsic electromagnetic fluctuations of a finite-temperature plasma at equilibrium. The two electron populations are uniformly distributed in space and have a Maxwell-Jüttner distribution function in momentum space, Eq.~\eqref{eq:Maxwell-Juttner}.
The implementation in the PIC code of the relativistic drifting Maxwell-Jüttner distribution follows the algorithm presented in Ref.~\cite{Zenitani2015}. 
Two series of simulations are carried out with temperature $T\simeq3.2\times 10^{-4} m_ec^2$ [correspondingly $T_L\simeq 10^{-4}(\gamma_0 -1)m_ec^2$ in the laboratory frame] and $T\simeq 0.1 m_ec^2$ [$T_L=3.3\times10^{-2}(\gamma_0 -1)m_ec^2$] referred to
in the following as quasi-cold case and warm case, respectively.

The length of the simulation box is $L_{x}\simeq50c / \omega_p$ and the cell extension is $\Delta x=\lambda_{D}/2$, where $\lambda_{D}$ is the Debye length $\lambda_{D}=\sqrt{T_L/(4\pi n_0 e^2)}$. The time resolution is $c\Delta t=0.95\Delta x$. The number of macro-particles-per-cell per species is $N_{p}=2000$. 

\begin{figure}[htb]
\centering
\includegraphics[width=0.45\textwidth]{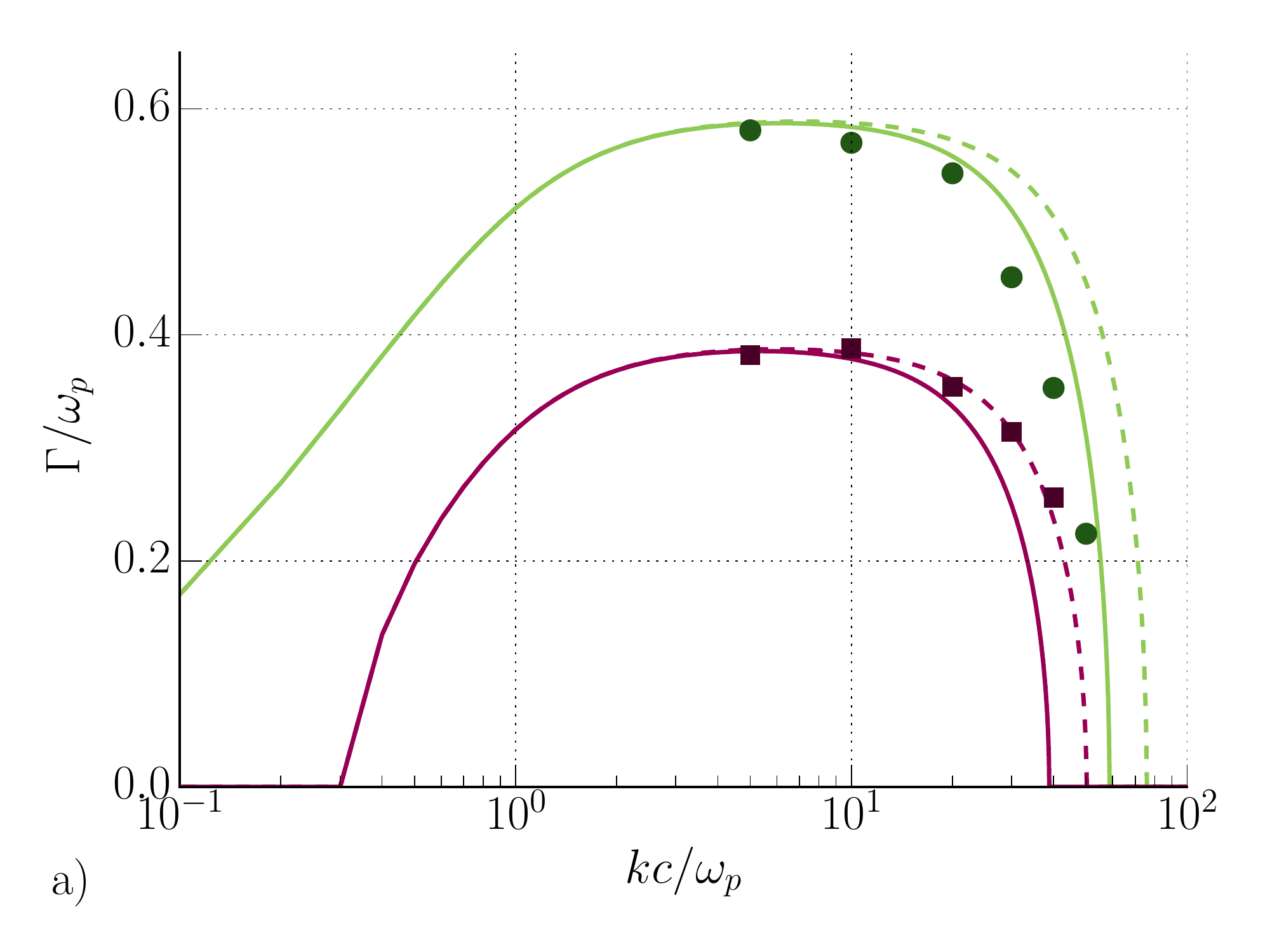}
\includegraphics[width=0.45\textwidth]{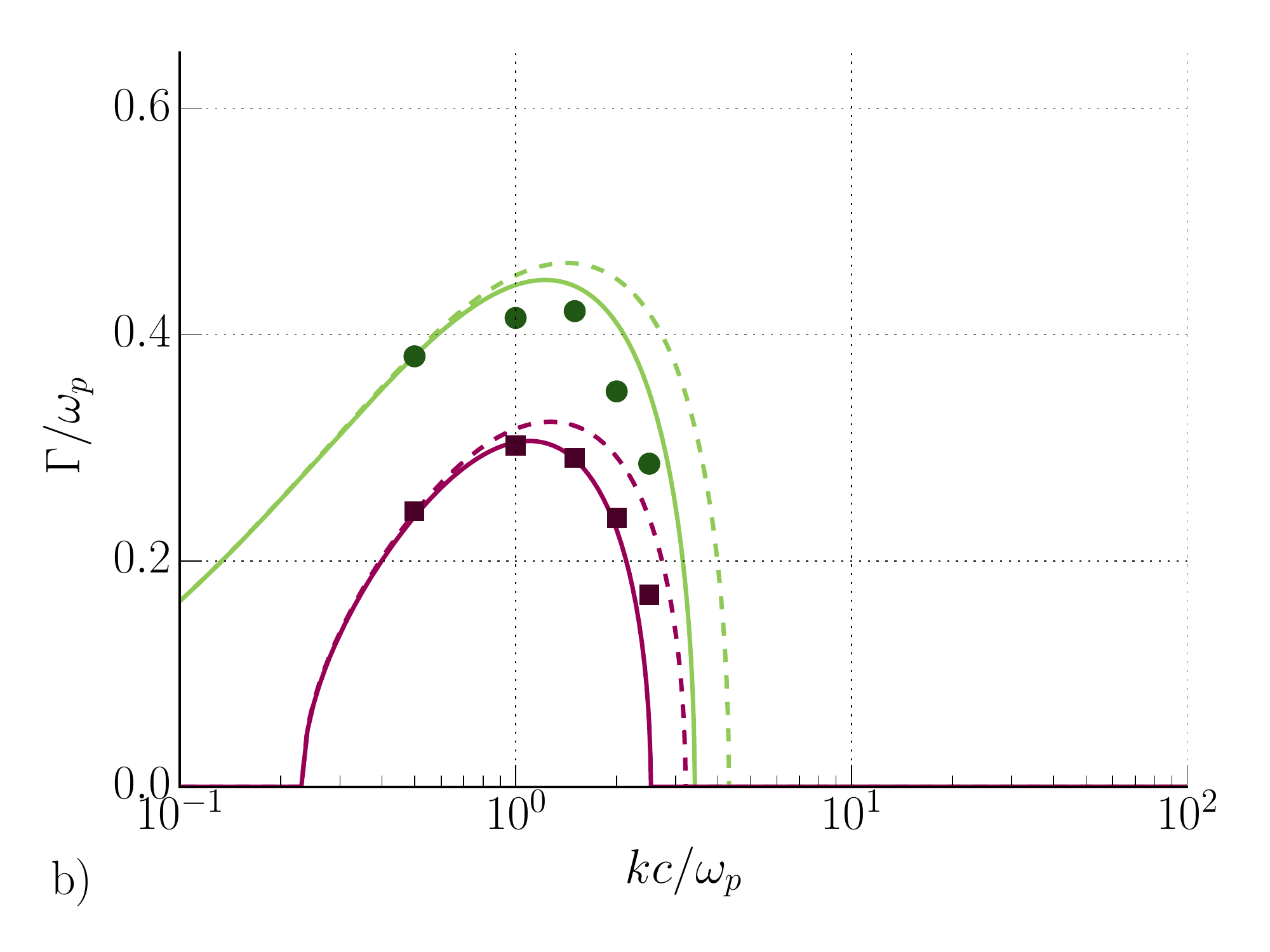}
\caption{\label{fig:GrowthRateTemperature}(color online) Growth rate of the instability as a function of the wavenumber. Theoretical predictions are computed from Eq.~\ref{eq:GrowthRateTemp} assuming isothermal closure (dashed lines) or adiabatic closure (plain lines). 
a) Quasi-cold $B_0=0$ (light green lines) and $B_0=0.75\,B_c$ (dark purple lines). 
b) Warm cases $B_0=0$ (light green lines) and $B_0=0.75\,B_c$ (dark purple lines). 
PIC simulations with $B_0=0$ (circles) and $B_0=0.75\,B_c$ (squares).
}
\end{figure}

The growth rate of different modes has been extrapolated from PIC simulations performing a Fourier analysis and measuring the growth of each mode independently. Results are reported in Fig.~\ref{fig:GrowthRateTemperature} for the two temperatures (quasi-cold and warm cases), with and without external magnetic field ($B_0=0$ and $B_0=0.75\,B_c$). Theoretical predictions from Eq.~\eqref{eq:GrowthRateTemp} are also shown considering both the adiabatic closure (solid lines) and isothermal closure (dashed lines).  
A fairly good agreement is found between PIC simulations and theory. The adiabatic assumption is consistent with all the simulation results except for the magnetized quasi-cold one, that more closely follows the isothermal curve. The problem of closure, kinetic effects and the limits of fluid approach are beyond the scope of this work and will be discussed elsewhere. Our results nevertheless suggest that the proposed relativistic fluid approach, which gives tractable solutions for the growth rate, is relevant to model the Weibel instability in the regimes discussed here.

\subsection{Nonlinear phase and saturation}\label{sec:SaturationMultiMode}

\begin{figure}[htb]
\centering
\includegraphics[width=0.45\textwidth]{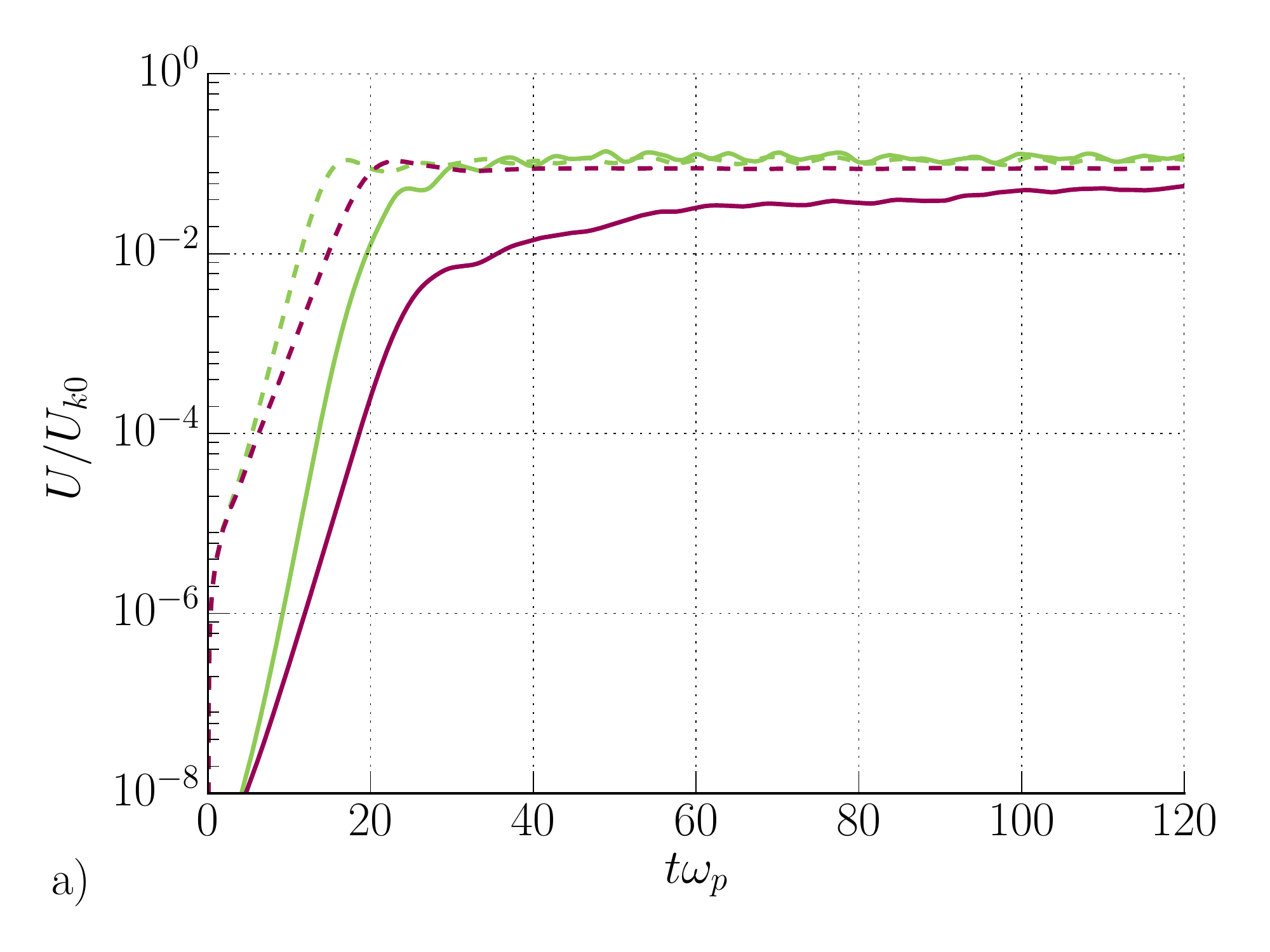}
\includegraphics[width=0.45\textwidth]{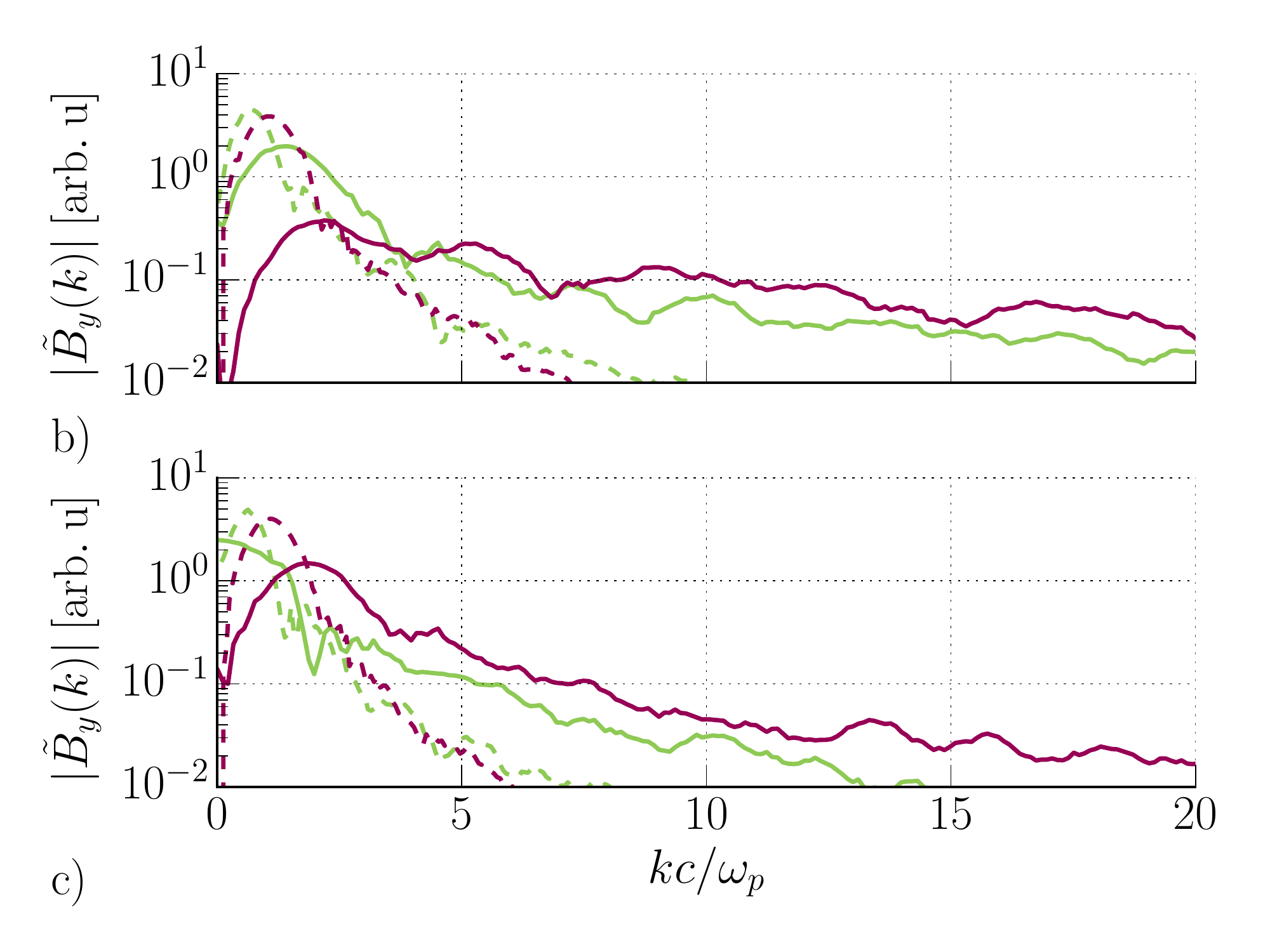}
\caption{\label{fig:EnergyT1T2}(color online) a) Evolution of the magnetic energy of the field $B_y$. b) Spectrum of $B_y$ at $t=30\,\omega_p^{-1}$. c) Spectrum of $B_y$ at the end of the simulations $t=120\,\omega_p^{-1}$.
Quasi-cold simulations (plain lines), warm simulations (dashed lines), with $B_{0}=0.0$ (light green lines) and $B_0=0.75\,B_c$ (dark purple lines). Spectra are shown after the application of Savitzky-Golay filter~\protect\cite{Filter} in $k-$space to reduce the noise. }
\end{figure}
Let us now study the non-linear phase and saturation of the instability.
Figure~\ref{fig:EnergyT1T2}a shows the evolution of the energy $U$ associated with the Weibel generated magnetic field $B_y$ normalized to the total initial flow energy $U_{k0}$. In the quasi-cold simulations (plain lines), the saturation level is modified by the application of the external field $B_0$. For the magnetized plasma (dark purple line), saturation is reached at $t_{\rm sat,B_0}\simeq 25\,\omega_p^{-1}$, identified by the clear change in the slope in Fig.~\ref{fig:EnergyT1T2}a. This stage corresponds to the saturation of the modes with large wavenumbers $k\geq 10\,\omega_p/c$, that grow with the largest rate Fig.~\ref{fig:GrowthRateTemperature}a and saturate with a low level of the magnetic field as predicted by the trapping mechanism Fig.~\ref{fig:SaturationMechanisms}a. Indeed, at that time, the amplitude of the oscillations of the magnetic field $B_y\simeq 0.12\,m_e\omega_pc/e$, is consistent with the saturation predicted for those modes. 
The saturation of these modes occurs at the same level of $U$ for the simulation with $B_0 = 0$ (light green line) around $t^*\simeq16\,\omega_p^{-1}$. Note that $t^*<t_{\rm sat,B_0}$ as expected due to the larger growth rate in the absence of an external magnetic field.

After this first saturation stage, the magnetized and unmagnetized cases evolve differently.
In Fig.~\ref{fig:EnergyT1T2}a the two plain lines do not reach the same level. Even at later time (not shown here) the slow rise in the magnetized curve ceases, the energy reaches an asymptotical value lower than the unmagnetized one and remains constant after $t\simeq300\,\omega_p^{-1}$.

In the unmagnetized plasma, once the large wavenumber modes have reached saturation, the small-$k$ modes keep growing up to their own saturation level. 
The growth of small $k$ filaments involves a rearrangement in large structures of the particles with opposite flow velocity.
In the magnetized case, in order to create filaments with small $k$, not only the currents should be redistributed but also the external magnetic field lines, that during the linear phase are compressed inside the filaments. 

This process entails a slowdown in the growth of small-$k$ modes, hence the very low slope in Fig.~\ref{fig:EnergyT1T2}a. Thus, with the introduction of the external magnetic field the large-$k$ modes remain stable after their saturation and this affects the growth of the modes not yet saturated. This is clearly shown in the spectra of the magnetic field $B_y$ reported in Fig.~\ref{fig:EnergyT1T2}b for all four simulations at $t=30\,\omega_p^{-1}$ and at $t=120\,\omega_p^{-1}$. In the quasi-cold magnetized case the spectrum is dominated by the large-$k$ modes, while in the unmagnetized case there is a dominant mode with $k\simeq 1.2\,\omega_p/c$. 

The increase of the initial temperature limits the range of unstable wavenumbers due to the temperature effect of stabilizing the large-$k$ modes, Fig.~\ref{fig:GrowthRateTemperature}b. In this way, the saturation level becomes again independent from the external magnetic field, Fig.~\ref{fig:EnergyT1T2}a (dashed lines). The spectrum of $B_y$ at the saturation is peaked around $k\simeq0.7\,\omega_p/c$ for the unmagnetized case (dashed light green line) and $k\simeq0.9\,\omega_p/c$ for the magnetized case (dashed dark purple line). The peak values are in good agreement with the $k$ predicted to have the highest saturation level in the cold single-mode simulations $k=0.86\,\omega_p/c$, Fig.~\ref{fig:SaturationMechanisms}a.

To summarize, the saturation level at large temperature does not depend on the application of the external flow-aligned magnetic field and the spectra are peaked around the optimal value found in the cold case, while at low temperature 
the energy transfer towards small-$k$ filaments is hampered by the magnetic field, resulting in a lower saturation amplitude and a wider distribution in $k$. 

\subsection{Late merging phase}\label{Merging}

At later times, after the saturation of the instability $t>30\,\omega_p^{-1}$, the so-called merging or coalescence of filaments governs the dynamics of the system. During this phase the total energy in the magnetic field remains roughly constant, see Fig.~\ref{fig:EnergyT1T2}a. The merging of two filaments is the result of the attractive force between filaments of parallel current. Regarding the spectrum of the Weibel generated magnetic field, the coalescence of filaments involves a shift toward small wavenumber modes as it creates structures of increased transverse size in the current and accordingly in the magnetic field.
Simplified models for the coalescence of filaments in cylindrical geometry have been presented in Refs.~\cite{AchterbergNL,Medvedev2005}.  
In our 1D geometry, the merging of filaments could be quite unexpected. Indeed, in order to observe the coalescence, the attractive force between two filaments of parallel current should overcome the repulsive force due to the filament of opposite current in the middle of them. Thus, a series of equal positive and negative current filaments would produce a stable situation, the attractive and repulsive force balancing each other, as observed in single mode simulations. However, in the case of a broad spec aretrum of unstable modes, merging can occur as this balance is not achieved due to (i)~intrinsic irregularity in the filament spatial distribution and (ii) the effect of the inductive electric field, as detailed below. 

\begin{figure}
\centering
\includegraphics[width=0.5\textwidth]{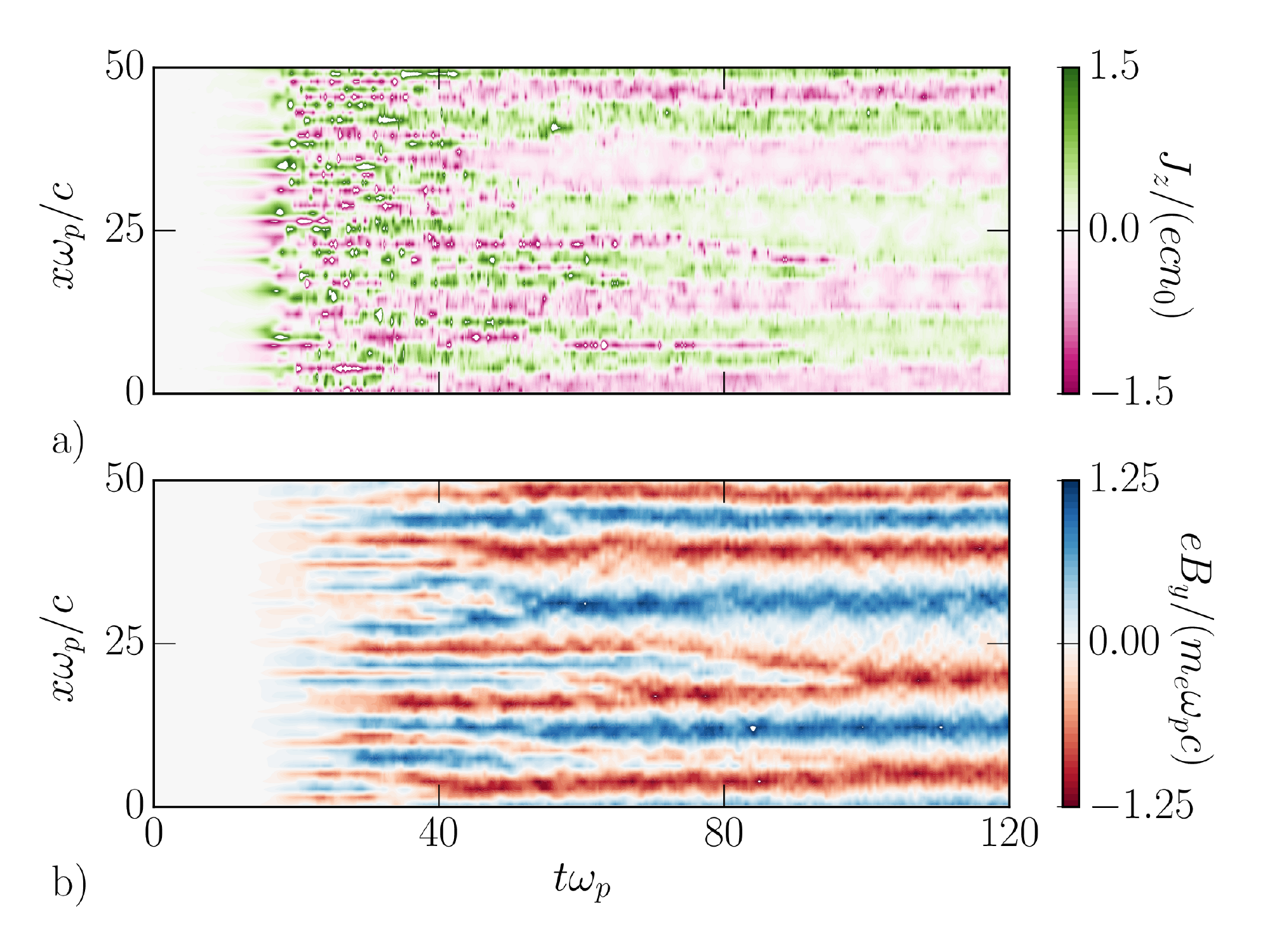}
\caption{\label{fig:CurrentBy_mm}(color online) a) Evolution of the current $J_z$ of the two counter-streaming beams. b) Evolution of the Weibel generated magnetic field $B_y$, for the unmagnetized quasi-cold case.}
\end{figure}
\begin{figure}
\centering
\includegraphics[width=0.45\textwidth]{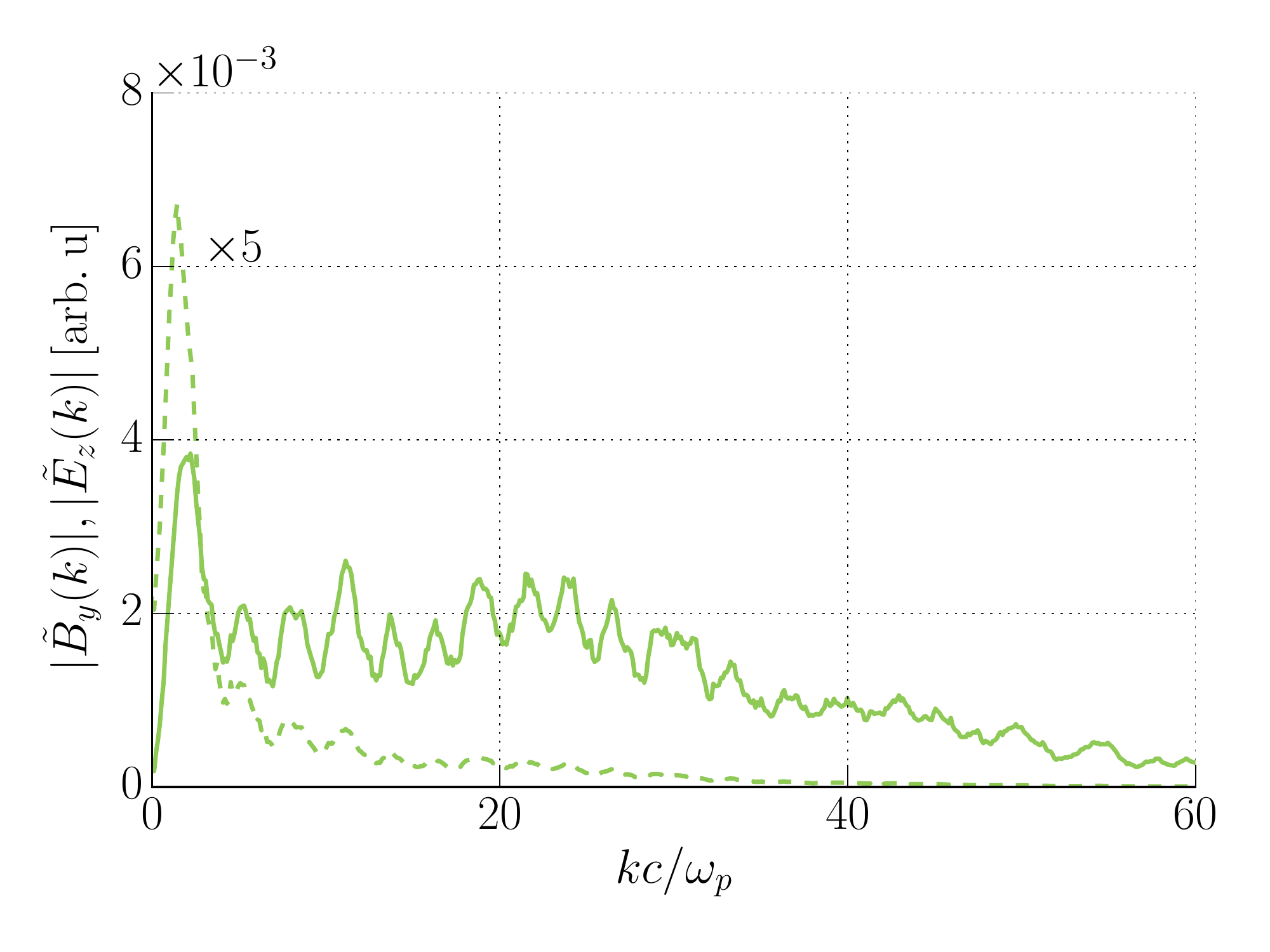}
\caption{\label{fig:ByEz_k_End}(color online) Spectrum of $B_y$ (plain line) and $E_z$ (dashed line) in the linear phase $t=10\,\omega_p^{-1}$, for the unmagnetized quasi-cold case. Spectra are shown after the application of Savitzky-Golay filter~\protect\cite{Filter} in $k$-space to reduce the noise. }
\end{figure}

In Figure~\ref{fig:CurrentBy_mm} the evolution in time of the current $J_z$ of the two counter-streaming electron beams and the Weibel generated magnetic field are shown for the unmagnetized quasi-cold simulation. 
At time $t=0$ the total current vanishes, then the filaments start to develop and the saturation is reached at $t\simeq 30\,\omega_p^{-1}$. The magnetic energy in Fig.~\ref{fig:EnergyT1T2}a remains constant after saturation and events of coalescence are clearly shown in Fig.~\ref{fig:CurrentBy_mm}.

Coalescence processes, even if in a 1D configuration,
can be explained as follows. In this series of simulations, the instability starts from a broad spectrum of modes. As already pointed out this leads to an intrinsic irregularity (randomness) in the filament spatial distribution. Furthermore this entails a difference in the spectrum of the Weibel-generated magnetic field $B_y$ and the spectrum of inductive electric field $E_z$, as shown in Fig.~\ref{fig:ByEz_k_End}. This difference in $k$-space can be explained considering a broad spectrum in the magnetic field $B_y$ in the linear phase of the instability 
\begin{eqnarray}\label{eq:MergingTempBy}
B_y(x,t)&=&\sum_n \tilde{B}_{y0,n}\sin(k_n x)e^{\Gamma(k_n)t} 
\end{eqnarray} 
and using Maxwell-Faraday equation to compute the inductive electric field $E_z$
\begin{eqnarray}\label{eq:MergingTempEz}
E_z(x,t)&=&\sum_n \tilde{B}_{y0,n}\frac{\Gamma(k_n)}{ck_n}\cos(k_n x)e^{\Gamma(k_n)t}
\end{eqnarray} 
the sum running over all wavenumbers. We assume the same amplitude for each mode at early time, so that $\tilde{B}_{y0,n}$ is independent from $k_n$. 
Due to the factor $\Gamma(k)/k$, considering $\Gamma(k)$ as calculated from Eq.~\eqref{eq:GrowthRateTemp}, the inductive electric field $E_z$ vanishes at large $k$, so that its spectrum peaks at small $k$. 

Despite the amplitude of $E_{z}$ is smaller than $B_{y}$, it can play a key role due to the the different spectrum with respect to $B_y$.
Since the electric field has a peak in the spectrum at small $k$, i.e. large wavelengths, it can have opposite effect on two neighbor filaments with opposite current, corresponding to a mode with large $k$. $E_z$ accelerates the particles of one filament while decelerating the other. The unbalance produced in the current allows for the merging of the filaments. The attractive force between two filaments of positive currents, whose particles are accelerated by $E_z$, exceeds the repulsive force due to the negative filament in the middle, for which $E_{z}$ is decelerating, resulting in the coalescence of the positive currents. 
In the simulations with single seeded mode, $E_z$ and $B_y$ have the same periodicity, see Eqs.~\eqref{eq:BlinTheory} and~\eqref{eq:ElinTheory}, so that $E_z$ tends to slow the electrons of both the counter-streaming beams down. The filaments form a regular structure of identical positive and negative filaments, and merging is not be observed.

At saturation, in all simulations, except in the magnetized quasi-cold case, the spectra of the magnetic field $B_y$ have a peak for $k\lesssim \omega_p/c$, as shown in Fig.~\ref{fig:EnergyT1T2}b. The corresponding spectra at the end of the simulation $t=120\,\omega_p^{-1}$, show that the peak is increased, narrower and slightly shifted toward a lower $k$. After the saturation the energy in the magnetic field is constant, Fig.~\ref{fig:EnergyT1T2}a, thus the evolution of the peak is a signature of the merging events, that transfer energy to the modes with large wavelengths. 
In the quasi-cold simulation, the presence of the external magnetic field produces a broad spectrum of modes at the saturation, that remains much broader than in the other cases, also at the end of the simulation. The coalescence of filaments is hampered by the external magnetic field, since an additional energy is required to move the magnetic field lines.

\section{Conclusions}\label{Conclusions}

The Weibel instability driven by two symmetric counter-streaming relativistic electron beams in the presence of a flow-aligned magnetic field has been investigated using both analytic modeling and 1D3V PIC simulations.

The linear stage of the instability is modeled using a relativistic fluid approach accounting for the effect of the electron pressure in the case of finite temperature plasma flows.
This fluid model gives tractable solutions for the growth rate which are found to be in good agreement with the PIC simulations. 

The saturation (nonlinear phase) of the instability has then been investigated. Considering a single growing mode, the mechanisms responsible for saturation in the presence of the external magnetic field have been clarified. At small wavenumber the dominant role of the Alfvén current-limitation is highlighted. We show that the external magnetic field can slightly increase the field amplitude at saturation. In the large wavenumber limit, the trapping mechanism leads to the saturation of the instability. The predicted saturation level is found to be independent of the strength of the external magnetic field, as long as the latter remains smaller than the well-known critical field above which the instability is quenched. These theoretical results are in good agreement with PIC simulations seeded with a single mode. 

The saturation and late merging stages have also been investigated in PIC simulations with the instability seeded from broadband thermal fluctuations.
In a low temperature plasma, the average saturation level is decreased by the application of an external magnetic field, since after the saturation of large-$k$ modes 
the external magnetic field hinders the redistribution of energy towards small $k$. Even at late times, the Weibel magnetic field spectrum in a magnetized plasma is much broader then in the unmagnetized case. As a consequence, filament merging is also inhibited by the external magnetic field. Increasing the initial flow temperature, the saturation level is found to be independent from the external magnetic field, and the Weibel field spectra are found to be peaked around an optimal wavenumber which value is well predicted considering the saturation mechanisms using a single mode analysis. 

Merging processes have been identified in our 1D simulations. The mechanisms that allow for this merging in multi-mode simulations and not in single-mode have been explained as following from both, the irregular distribution of filaments growing from thermal fluctuations and the effect of the small $k$ inductive field.\\

The analysis of the growth and saturation of the Weibel instability performed in this paper allow for a generalization of the saturation mechanisms in presence of a magnetic field aligned with the flows and shows that the saturation stage is only weakly affected.

Our results can be applied to astrophysical systems where the Weibel instability is driven in magnetized plasmas (i.e. pulsar wind) and that are related to collisionless shock formation and particle acceleration.  
In the context of laser-plasma interaction, and of relativistic laboratory astrophysics in particular, these results suggest that using a guiding external magnetic will not strongly modify the level of Weibel-generated magnetic fields while helping maintain a high plasma density, hence fastening the development of plasma instabilities and potentially of the formation of collisionless shocks.

\section*{Acknowledgements}
The authors thank Laurent Gremillet for fruitful discussions and the SMILEI development team for technical support. AG and CR acknowledge financial support from Grant No. ANR-11-IDEX-0004-02 Plas@Par. AG acknowledges support from the Université Franco-Italienne through the Vinci program. 
This work was performed using HPC resources from GENCI-TGCC (Grant 2016 - x2016057678).

\appendix
\section{Alfvèn limit in the presence of an external (guiding) magnetic field}\label{AppendixAlfven}

The Alfvèn limit defines the maximum (critical) current that a beam of charged particles can sustain before the particle trajectories, in the self-generated magnetic field, start limiting the current itself due to the reduction and/or inversion of particle motion in the flow direction~\cite{Alfven1939}. The maximum current can be defined in different ways~\cite{Alfven1939,HammerRostoker1970,Honda2000} that, within a factor, give very similar results. In this Appendix, we follow more closely the original approach proposed by Alfvén.

The derivation presented here considers a 1D3V geometry and given sine-profile for the current density to be consistent with our PIC simulations.
Generation to the more realistic 2D(r,z) geometry and arbitrary profile is straightforward. 
The critical current for a uniform cylindrical current in 2D(r,z) geometry is given at the end of this Section.

Let us start by assuming a sinusoidal profile for the current density $J_z$ for $-\pi/2 \le kx \le\pi/2$
\begin{eqnarray}\label{eq:CurrentProfile} 
J_z(x) = -J_0\cos(kx)\,
\end{eqnarray}
and consistently with Ampère's law the magnetic field
\begin{eqnarray}\label{eq:ByProfile}
B_y(x) = -2\pi I_{\rm 1D}^{(0)}/c\,\sin(kx)\, 
\end{eqnarray}
with $I_{\rm 1D}^{(0)}=2\,J_0/k$ the absolute value of the total (areal) current.

Considering this magnetic field and the external (guiding) magnetic field ${\bf B}_0=B_0 \hat{\bf z}$ as time-independent, three constants of motion allow for the description of an electron dynamics in these fields: the electron energy (Hamiltonian) $\mathcal{H}=m_e c^2\,\sqrt{1+{\bf p}/(m_e^2c^2)}$, and the two components of the electron canonical momentum ${\bf \Pi}={\bf p}-e{\bf A}/c$ lying in the $(y,z)$-plane. The vector potential ${\bf A}$ is computed inverting the relation ${\bf B}={\bf \nabla}\times {\bf A}$, leading to
\begin{eqnarray}
\label{eq:Ay}  A_y(x) &=& B_0 x  \\
\label{eq:Az}  A_z(x) &=& 2\pi I_{\rm 1D}^{(0)}/(kc) \left[1-\cos(kx)\right]
\end{eqnarray}
where we have taken $A_y(0)=A_z(0)=0$. Considering an electron initially located at the border of the filament $x_0=\pi/(2k)$ with initial momentum $m_e\gamma_0 v_0 \hat{\bf{z}}$ (correspondingly $\mathcal{H}_0=\gamma_0 m_e c^2$), one gets:
\begin{eqnarray}
\label{eq:H0} p_x^2 &=& m_e^2c^2 (\gamma_0^2-1)-p_y^2-p_z^2\\
\label{eq:py} p_y &=& -e B_0 x_0/c\,(1-x/x_0)\\
\label{eq:pz} p_z &=& m_e \gamma_0 v_0 - 2\pi e I_{\rm 1D}^{(0)}/(kc^2)\,\cos(kx)\,.
\end{eqnarray}
The critical current $I_{\rm 1D}^{(c)}$ is then defined as the minimum current for which the longitudinal momentum Eq.~\eqref{eq:pz} vanishes leading to:
\begin{eqnarray}\label{eq:CritCurr1DGen}
I_{\rm 1D}^{(c)} = \frac{1}{2\pi}\frac{m_e c^2}{e}{\rm min}\left\{\frac{\gamma_0 v_0 k}{\cos(kx)}\right\}\,.
\end{eqnarray}

In the absence of external magnetic field ($B_0=0$, and $p_y=0$ at all times), this minimum is reached for $x=0$, i.e. when the electron longitudinal momentum vanishes on-axis, leading to:
\begin{eqnarray}\label{eq:CritCurr1D}
I_{\rm 1D}^{(c)} = \frac{1}{2\pi}\frac{m_e c^2}{e}\,\gamma_0 v_0 k\,,
\end{eqnarray}
that corresponds to the magnetic field strength [$B_y$ from Eq.~\eqref{eq:ByProfile}]
\begin{eqnarray}
B_{\rm max} = \gamma_0 v_0 k\,\frac{m_e c}{e}\,,
\end{eqnarray}
given by Eq.~\eqref{eq:Bsat_smallk}.

In the presence of a guiding magnetic field ($B_0 \ne 0$), the electron starting at the border $x_0$ of the current may not reach its center $x=0$ before being turned back under the effect of the guiding magnetic field.
As a consequence, the critical current Eq.~\eqref{eq:CritCurr1DGen} has to be computed taking $x=x^{*}$ with $x^{*} = 0$  if the electron can reach the center of the current,
and $x^{*}>0$ the turning point of the electron when this one cannot reach $x=0$. For large enough external magnetic field [$B_0>(\gamma_0 v_0/x_0) \,m_e c/e$, correspondingly $A \equiv v_0 /\vert\Omega_0 x_0\vert<1$], one obtains $x^{*}$ as the point for which $p_x=p_z=0$ (all the electron momentum is in $p_y$) leading to $x^{*}=x_0\,(1-A)$.
For lower value of the external magnetic field ($A \ge 1$), the electron will eventually reach the center of the filament so that $x^{*}=0$. This leads to the critical current
\begin{eqnarray}\label{eq:CritCurr1DB0}
I_{\rm 1D}^{(c)} = \frac{1}{2\pi}f(A)\,\frac{m_e c^2}{e}\,\gamma_0 v_0 k\,,
\end{eqnarray}
with $f(A)=[\cos(\pi(1-A)/2)]^{-1}$ for $A<1$, and $f(A)=1$ otherwise,
corresponding to the magnetic field strength
\begin{eqnarray}
B_{\rm max} = f(A)\,\gamma_0 v_0 k\,\frac{m_e c}{e}\,,
\end{eqnarray}
given by Eq.~\eqref{eq:Bsat_smallkB0}.\\

A similar derivation can be done in the case of a uniform cylindrical current (with radius $R$) in 2D(r,z) geometry.
The constants of motions are then given by the Hamiltonian, $z$-component of the canonical momentum and canonical angular momentum.
One then obtains the critical current:
\begin{eqnarray}\label{eq:CritCurr}
I^{(c)} = I_0\,\frac{\gamma_0 v_0/c}{1-r^{*}/R}\,,
\end{eqnarray}
with $I_0 = m_e c^3/e \simeq 17~{\rm kA}$, and for which $r^{*}$ plays the same role as $x^{*}$ in 1D3V and depends on the external magnetic field as
\begin{eqnarray}
r^{*} = \frac{R}{2}\,\left(\sqrt{4+A^2}-A\right)\,,
\end{eqnarray}
with $A=v_0/\vert \Omega_0 R\vert$. In the absence of external magnetic field $A\rightarrow \infty$ ($r^{*}=0$), one recovers the well-known result by Alfvén.

Notice that both Eqs.~\eqref{eq:CritCurr1DB0} and~\eqref{eq:CritCurr} predict an increase of the critical current with the application of a guiding magnetic field.
The possibility to exceed the Alfvén limit by applying an external magnetic field along the flow direction was already considered, e.g. in Ref.~\cite{Peratt} Par.2.5.2-6.

\end{document}